%% file: ms.tex
\documentclass[preprint,12pt]{aastex}

\shorttitle{BHR7 Envelope and Disk}
\shortauthors{Tobin et al.}

\newcommand{\nthp}{\mbox{N$_2$H$^+$}}

\newcommand{\ntdp}{\mbox{N$_2$D$^+$}}

\newcommand{\cateo}{\mbox{C$^{18}$O}}
\newcommand{\thco}{\mbox{$^{13}$CO}}

\newcommand{\kms}{\mbox{km s$^{-1}$}}

\begin{document}

\title{The Envelope Kinematics and a Possible Disk Around the Class 0 Protostar within BHR7}
\author{John J. Tobin\altaffilmark{1,2}, Steven P. Bos\altaffilmark{2}, Michael M. Dunham\altaffilmark{3,4}, Tyler L. Bourke\altaffilmark{5}, Nienke van der Marel\altaffilmark{6}}

\begin{abstract}
We present a characterization of the protostar embedded within the BHR7 dark 
cloud, based on both photometric measurements
from the near-infrared to millimeter and interferometric continuum 
and molecular line observations at millimeter wavelengths. We find that
this protostar is a Class 0 system, the youngest class of protostars, measuring its bolometric
temperature to be 50.5~K, with a bolometric luminosity of 9.3~L$_{\sun}$.
The near-infrared and \textit{Spitzer} imaging show
a prominent dark lane from dust extinction separating clear bipolar outflow cavities. 
Observations of $^{13}$CO ($J=2\rightarrow1$), C$^{18}$O ($J=2\rightarrow1$), and other 
molecular lines with the Submillimeter Array (SMA) exhibit a clear rotation signature on scales $<$1300~AU.
The rotation can be traced to an inner radius of $\sim$170~AU and the rotation curve is consistent with
an R$^{-1}$ profile, implying that angular momentum is being conserved. 
Observations of the 1.3~mm dust continuum with the SMA reveal a resolved 
continuum source, extended in the direction
of the dark lane, orthogonal to the outflow. The deconvolved size of the continuum
indicates a radius of $\sim$100~AU for the continuum source at the 
assumed distance of 400~pc. The visibility amplitude profile of the continuum emission
cannot be reproduced by an envelope alone and needs a compact component.
Thus, we posit that the resolved continuum source could be tracing a Keplerian disk
in this very young system. If we assume that the continuum radius traces
a Keplerian disk (R$\sim$120~AU) the observed rotation profile is
consistent with a protostar mass of 1.0~$M_{\sun}$. 
\end{abstract}

\altaffiltext{1}{Homer L. Dodge Department of Physics and Astronomy, University of Oklahoma, 440 W. Brooks Street, Norman, OK 73019, USA}
\altaffiltext{2}{Leiden Observatory, Leiden University, P.O. Box 9513, 2300-RA Leiden, The Netherlands}
\altaffiltext{3}{Department of Physics, State University of New York Fredonia, Fredonia, New York 14063, USA}
\altaffiltext{4}{Harvard-Smithsonian Center for Astrophysics, 60 Garden St., Cambridge, MA, USA}
\altaffiltext{5}{SKA Organization, Jodrell Bank Observatory, Lower Withington, Macclesfield, Cheshire SK11 9DL, UK}
\altaffiltext{6}{Institute for Astronomy, University of Hawaii, 2680 Woodlawn Drive, 96822 Honolulu, HI, USA}
\altaffiltext{7}{Herzberg Astronomy \& Astrophysics Programs, National Research Council of Canada, 5017 West Saanich Road, Victoria, BC, Canada V9E 2E7}

\section{Introduction}

The formation of rotationally-supported disks during star formation is a crucial part of the
star and planet formation process. Once a disk has formed, stellar mass assembly will then be governed by 
accretion through
the disk, and the disk will further enable the growth of solids, catalyzing the planet formation 
process \citep{testi2014}. Furthermore, if the forming disks are sufficiently massive they 
may fragment via gravitational instability, leading to the formation of a close binary star 
system or possibly giant planets \citep[e.g.,][]{kratter2010,tobin2016}.
Also, clumps formed via gravitational instability may
lead to episodic bursts of accretion \citep{vorobyov2006}, and 
they may lead to disk snowline radii moving in and 
out during the protostellar phase \citep{cieza2016}. However, it is uncertain if (or how often) processes
such as gravitational instability can function in protostellar disks because their general properties
such as radii, mass, and temperature structure have yet to be 
characterized for a broad sample. This is because the continuum and 
molecular line emission of the envelope is entangled with that of the disk early in the star formation 
process, the Class 0 phase \citep{andre1993, dunham2014}, when the infalling envelopes are very dense.

In order to understand the formation and evolution of disks, they must be observed throughout the star formation process. Thus, both high spatial and spectral resolution
must be coupled with high-sensitivity observations to distinguish the disk
from the infalling envelopes. A few examples of disks in Class 0 systems
have been found \citep{tobin2012,murillo2013,ohashi2014,codella2014b,lee2014,lindberg2014,yen2015,yen2017}, having
a wide variety of sizes, from $>$100~AU to $<$10~AU. However, the total number of systems is still quite small,
therefore detections and/or evidence of disks in additional systems are important.

BHR7 \citep{bourke1995a} is an isolated dark cloud harboring an embedded protostar, discovered by
the coincidence 
of its mid to far-infrared emission detected by the 
Infrared Astronomy Satellite (IRAS 08124-3422), an optical 
dark cloud \citep{hartley1986}, and ammonia emission \citep{bourke1995b}. The isolated nature
of BHR7 makes it an ideal target to examine, free of confusion from 
the superposition of neighboring protostars. However, a drawback of an isolated system is its uncertain distance,
because of the difficulty to relate BHR7 to a particular star forming region. BHR7, however, is associated with the Vela Cometary globules and the distance estimates range between 400~pc to
500~pc \citep{woermann2001}. The systemic velocity of BHR7 indicates that it is on the near-side of an expanding shell, 
thus we adopt $\sim$400~pc as the distance to BHR7 throughout the text.
In order to distinguish between
the protostar and the dark cloud, we will refer to the protostar as BHR7-MMS and refer to the 
region around the protostar as BHR7.

Since the original detection, this protostar has received little attention in the form of follow-up observations.
However, the Two-Micron All Sky Survey \cite[2MASS; ][]{skrutskie2006} and \citet{santos1998} observed BHR7 and 
detected evidence of a bipolar scattered light nebula at 2.15~\micron, indicating the possible presence of 
outflow activity from this source. The bipolar
nebula also appeared as if it might be near edge-on in both sets of near-infrared observations
from the dark lane that appears to bisect the nebula.
Edge-on protostars systems 
can be advantageous because the orientation enables a clear view of the kinematics, 
well-separated from the outflow, and any velocity corrections
due to inclination are inherently small.

Here we present a multi-wavelength and multi-scale examination of BHR7 using ground and space-based
observations spanning the near-infrared to millimeter wavelengths, confirming the evolutionary
state of BHR7-MMS as a Class 0 protostar. We subsequently obtained observations with 
the Submillimeter Array \citep[SMA;][]{ho2004} in 
multiple configurations, gaining both a small-scale (sub-arcsecond) view, and a larger-scale (several arcsecond)
view of the kinematic structure. The key advance is that the SMA observations have enabled us to assess the likelihood of a disk in this protostar
system. The paper is structured as follows: the observations and data reduction are presented in Section 2, we discuss the
overall infrared imaging, spectral energy distribution (SED), and results from the SMA observations in Section 3, 
discussion of the results is presented in Section 5, and the Summary and Conclusions are given
in Section 5.

\section{Observations and Data Reduction}

\subsection{SMA Observations}
The SMA is located on Mauna Kea on the island of Hawaii at an altitude of 4000~m, and it nominally has eight 6~meter
antennas combined to form the interferometer. BHR7-MMS was observed by the SMA several times during 2015 and 2016 in
Very Extended, Extended, and Compact configuration at 1.3~mm; the observations are summarized in Table 1 and detailed in the following sections.

\subsubsection{Very Extended Observations}

The Very Extended observations were conducted on 2015 January 28 with 6 antennas operating and a maximum baseline of 362~m. The zenith atmospheric 
optical depth at 225~GHz was $\sim$0.06 throughout the observation. The observation used 
the ASIC correlator configured for continuum observations, sampling
a total bandwidth of 8~GHz (upper and lower sided bands). The spectral resolution was set to 64 channels per
104~MHz spectral chunk and the central frequency was 225.5 GHz. The observations used 
Callisto as the absolute flux calibrator, 3C279 as the bandpass calibrator, and 0747-331 
was used as the complex gain calibrator. The observations were conducted in a standard loop observing
the gain calibrator for 3 minutes and BHR7-MMS for 10 minutes. The total time on source was approximately 4.0~hrs.

\subsubsection{Compact Array Observations}

The Compact array observations were conducted on 2015 December 25 with 8 antennas operating and a maximum baseline of 69~m. The zenith atmospheric optical depth
at 225~GHz was $\sim$0.07 throughout the observation. For these observations, Uranus was used as the absolute flux calibrator,
3C84 was the bandpass calibrator, and 0747-331 was the complex gain calibrator.
The observations were conducted in a standard loop observing the gain calibrator for 3 minutes and BHR7-MMS for 15 minutes. 
The total time on source was approximately 5.0 hrs.
The correlator was configured for both spectral line and continuum observations with a central frequency of 225.4 GHz; 
we observed $^{12}$CO, $^{13}$CO, C$^{18}$O ($J=2\rightarrow1$), 
H$_2$CO ($J=3_{0,3}\rightarrow2_{0,2}$), and N$_2$D$^+$ ($J=3\rightarrow2$). All lines had a spectral resolution of 200~kHz (0.26~\kms), except C$^{18}$O and \ntdp\ which had a spectral resolution of 100 kHz ($\sim$0.13~\kms).
The total continuum bandwidth was 5.2~GHz (upper and lower sidebands combined).

\subsubsection{Extended Array Observations}

The Extended array observations were conducted on 2016 April 02 with 8 antennas operating and a maximum baseline of 181~m. The zenith atmospheric optical 
depth at 225~GHz was $\sim$0.04 throughout the observation. For these observations, Ganymede was used as 
the absolute flux calibrator, 3C273 was the bandpass calibrator, and 0747-331 was the complex gain calibrator.
The observations were conducted in a standard loop observing the gain calibrator for 3 minutes and BHR7-MMS for 15 minutes. 
The total time on source was approximately 4.2~hrs.
The correlator was configured for both spectral line and continuum observations; it has the same configuration and spectral resolution as
the Compact observations, except that H$_2$CO ($J=3_{2,1}\rightarrow2_{2,1}$), SO ($J_N=5_6\rightarrow4_5$), and SiO ($J=5\rightarrow4$) were also observed. They were not observed in the Compact observations due to a correlator problem. The SiO ($J=5\rightarrow4$) transition was also covered at the continuum spectral resolution (1.625 MHz or 2.25~\kms). The total
continuum bandwidth was 10.8~GHz (upper and lower sidebands combined; 5.2~GHz from the original correlator and 5.6 GHz from the early SWARM correlator).
 
\subsubsection{Data Reduction}

The SMA data were calibrated and edited using standard techniques within the MIR software package. MIR is an IDL-based
software package originally developed for the Owens Valley Radio, Observatory and adapted by the SMA group.
We performed subsequent imaging in MIRIAD \citep[Multichannel Image Reconstruction, Image Analysis and Display; ][]{sault1995} for both the spectral line and continuum data. However,
we did import the Very Extended data into CASA \citep{mcmullin2007} for self-calibration. We did this
by first exporting the continuum visibility data from MIR to MIRIAD format, then converting the MIRIAD visibility 
files to UVFITS via the MIRIAD task \textit{fits}, and then importing the data into CASA using the
\textit{importuvfits} task. The purpose of using CASA was to perform self-calibration on the continuum, using
the procedures established for VLA and ALMA data.
The final images from Very Extended were generated in CASA using the \textit{clean} task. Self-calibration of
the Very Extended data was useful given that the initial signal-to-noise (S/N) was 50 and phase-only self-calibration on 5 minute solution intervals increased the S/N to $>$100. The absolute flux density accuracy is between estimated to be between 10--20\%.

\subsection{Herschel Observations}

BHR7 was observed with the \textit{Herschel Space Observatory} as part of the 
Hi-Gal2$\pi$ project (OT2\_smolinar\_7), obsids 1342254511 and 1342254512, using the Spectral and Photometric Imaging Receiver (SPIRE) \citep{griffin2010}. While these data were acquired in a parallel mode
where SPIRE and the Photoconductor Array Camera and Spectrometer \citep[PACS;][]{poglitsch2010} observed simultaneously, the PACS coverage missed BHR7 by a few arcminutes. 
Therefore, BHR7 was
only observed by SPIRE at 250, 350, and 500~\micron. We used
the Jscan imaging products from the archive for photometry. We measured the flux density in 
a circular aperture around BHR7 with a radius of 40\arcsec\ and subtracted the background measured from a nearby region of blank sky. The background region was at RA=8:14:20.390, Dec=-34:36:26.29 and had
a radius of 129\arcsec. We used the median background value, and scaled this to the area of the 40\arcsec\
aperture. BHR7 is an isolated globule, so extended, diffuse emission from a surrounding 
cloud was not present, but there is an extended tail to the cloud, prohibiting the use of a background annulus. 
We used the extended source photometry methods described in the SPIRE data handbook\footnote{http://herschel.esac.esa.int/Docs/SPIRE/spire\_handbook.pdf}, using the extended source 
color correction factor and aperture corrections. We use values corresponding to a modified blackbody
with $\beta$=1.5, and a temperature of 30~K; these values are reasonably representative of 
the SED of the protostar that peaks near 100~\micron. The \textit{Herschel} SPIRE photometry 
are listed in Table 2; the uncertainties given are statistical only, accounting for the RMS noise in the image and
the number of pixels summed to measure the flux density. 
The PACS and SPIRE absolute photometric accuracy are estimated to be 3-5\%\footnote{http://herschel.esac.esa.int/Docs/PACS/html/pacs\_om.html} 
and 4\%\footnote{http://herschel.esac.esa.int/Docs/SPIRE/html/spire\_om.html}, respectively. However, in the SPIRE
bands there is additional extended emission that we do not capture with the apertures used; thus, the submillimeter
flux densities may be systematically low.

\subsection{Spitzer Observations}

BHR7 was observed with the Infrared Array Camera \citep[IRAC;][]{fazio2004} onboard the \textit{Spitzer Space Telescope}
on 2008 December 22 as part of program 50477 (PI: Bourke) during the cryogenic portion of the mission. This program observed
BHR7 at 3.6, 4.5, 5.8, and 8.0~\micron. Using the pipeline-processed images
from the \textit{Spitzer} Heritage Archive, we measured the photometry in the IRAC bands using
a circular aperture radius of 25\arcsec\ ($\sim$10000 AU) and calculated the background using the median of an off-source patch of sky. The \textit{Spitzer} photometry are listed in Table 2.

\subsection{CTIO Near-Infrared Observations}

BHR7 was observed at Cerro-Tololo Inter-American Observatory (CTIO) on 2009 June 11 using the Infrared SidePort Imager
 \citep[ISPI; ][]{vanderbliek2003} on the Blanco 4 m telescope. The conditions during the observations
were photometric
and the seeing was $\sim$0\farcs9. 
ISPI provides a $\sim$10\arcmin\ field of view on
a 2048$\times$2048 detector array. We observed BHR7 in J (1.25~\micron), H (1.6~\micron), and Ks (2.15~\micron) bands
The observations were conducted in a 10-point box dither patter with a dither step
size of 60\arcsec. At each position in the dither, 4~$\times$~15s coadds were taken
in H and Ks-bands and 2~$\times$~30s coadds in J-band. The sky background was measured using a median combination of
the dithered frames; the step size of 60\arcsec\ preserved extended emission on this scale.
A total integration time on source was 40 minutes for J-band, 20 minutes for H-band, and 30 minutes for Ks-band.
Further details that discuss the combination of the individual frames to make the final mosaic are given in 
\citet{tobin2010}. The data were calibrated using the 2MASS photometry catalog \citep{skrutskie2006} and 
the photometry are listed in Table 2 and are extracted from the same aperture radius as the \textit{Spitzer} IRAC data.

\subsection{Swedish-ESO Submillimetre Telescope Observations}

The 1.2~mm (250 GHz) continuum observations were carried out with the 37
channel bolometer array SIMBA (SEST IMaging Bolometer Array) at the SEST
(Swedish-ESO Submillimetre Telescope) on La Silla, Chile, between June 2001
and August 2003, as part of a larger program to map the BHR \citep{bourke1995a} and Spitzer c2d
cores \citep{evans2003,kampgen2004}. The observations
were obtained in fast-scanning mode (80\arcsec\ per second), with multiple
maps obtained at different orientations (hour angles), to reduce artifacts
due to the mapping scheme.  To estimate the atmospheric opacity sky-dips
were undertaken about every three hours.  The resulting zenith opacity
values were 0.10-0.45.  The beam size of an individual SIMBA bolometer was
24\arcsec\ FWHM, and the positional accuracy was estimated to be
2-3\arcsec\ from frequent pointing scans on Centaurus A, $\eta$ Carina,
and Uranus. Flux calibration was determined to be accurate to about 10\%,
through maps of Uranus and Neptune.  The RMS at the map center is estimated
to be 18 mJy/beam.  All data were reduced and analyzed with the MOPSI
software according to the instructions of the SIMBA Observer's Handbook
(2002)\footnote{http://www.apex-telescope.org/sest/html/internal-access/OnlineManuals/Receivers/Simba/simba.ps}
and \citet{kampgen2002}.

\section{Results}
 The synthesis of the multi-wavelength dataset collected toward BHR7-MMS has enabled us to characterize 
the orientation, evolutionary state, luminosity, and the small-scale physical and kinematic structure. 
All these data together contribute to a much better understanding of this system as an important 
test case for star formation theories.

\subsection{Near-IR, \textit{Spitzer}, and \textit{Herschel} Imaging}

\begin{figure}
\begin{center}
\includegraphics[scale=0.45]{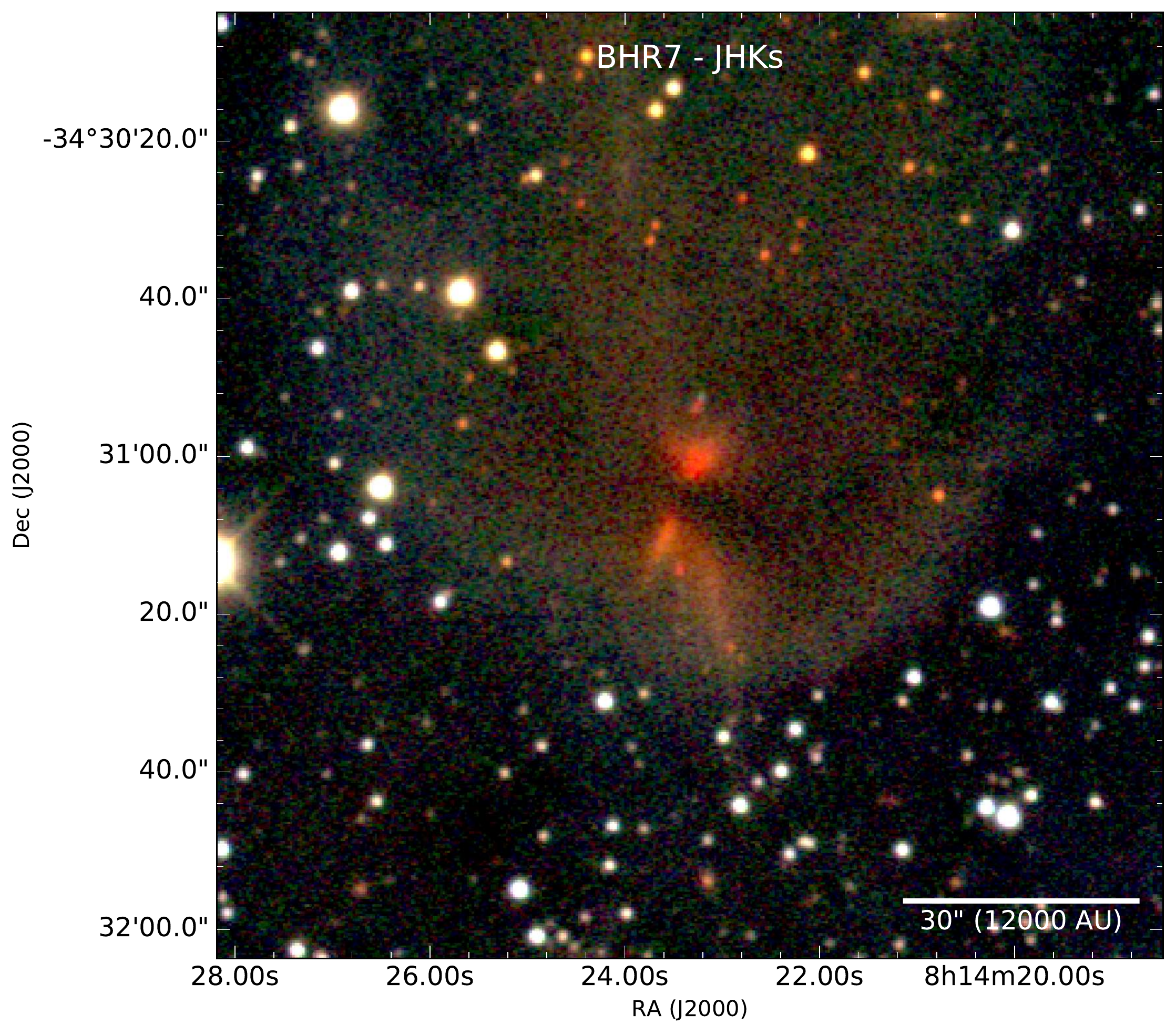}
\includegraphics[scale=0.45]{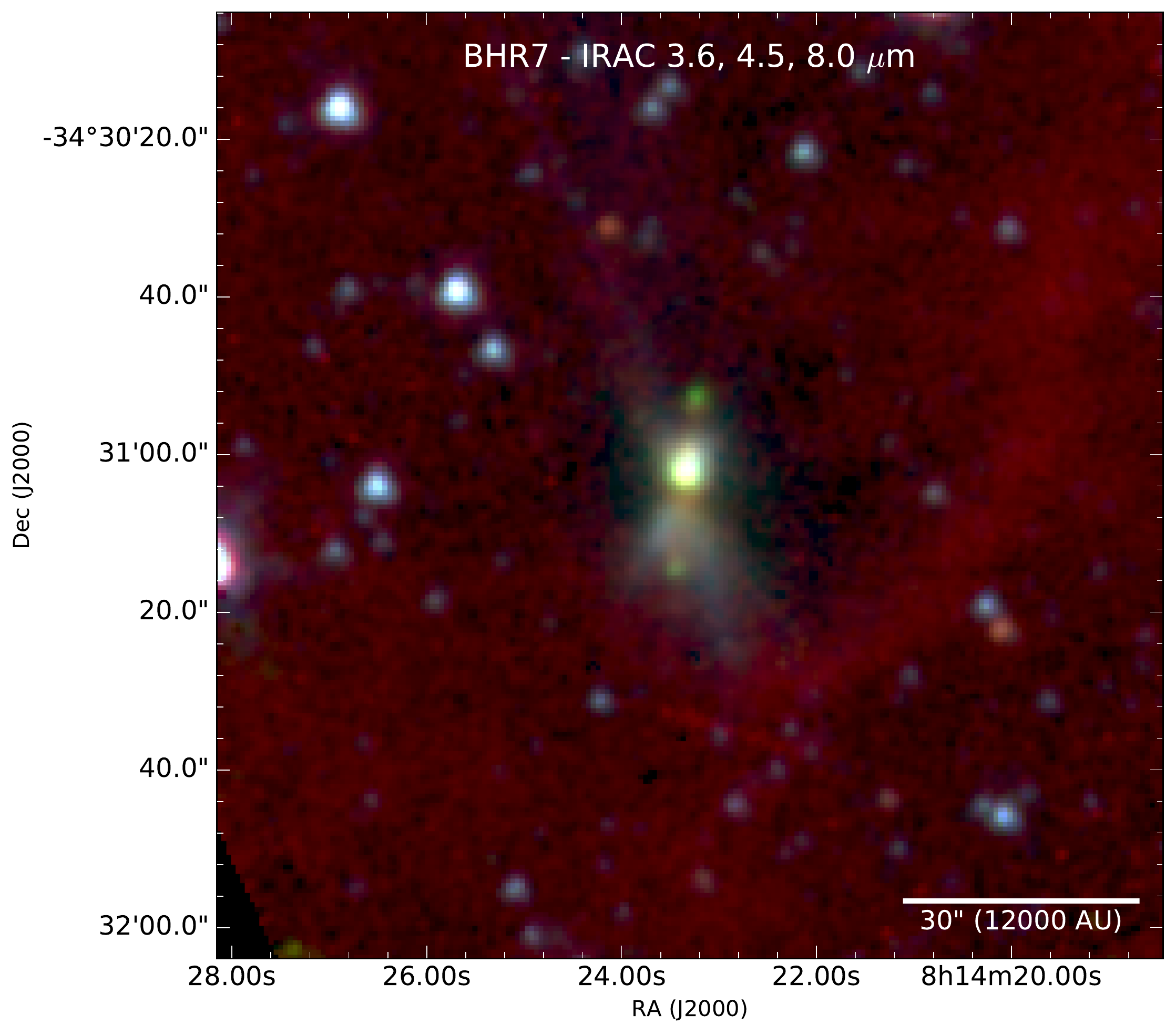}

\end{center}
\caption{Images of BHR7 at J, H, and Ks-bands (top) and 
\textit{Spitzer} IRAC 3.6, 4.5, 8.0~\micron\ (bottom);
the wavelength order for both images are assigned to 
blue, green, and red colors, respectively. Notice the thick dark lane separating
the north and south sides of the outflow cavity in both images. The outflow cavity
appears in scattered light in nearly the north-south direction. There are
also two knots in the center of the outflow cavity that are prominent in Ks-band
and at 4.5~\micron, tracing a shock in the outflow. The dark lane
appears thicker at JHKs-bands due to greater levels of extinction
and it also has a slightly different angle. There also appears
to be an `edge' to the envelope surrounding BHR7, appearing
in diffuse scattered light at JHKs-bands, but also in 8~\micron, likely reflecting
PAH emission.}
\label{bhr7-nir-irac}
\end{figure}

The near-infrared image with sub-arcsecond seeing and the \textit{Spitzer} IRAC
imaging shown in Figure \ref{bhr7-nir-irac} reveal the appearance of a 
prominent bipolar outflow cavity viewed in scattered
light from a central illuminating source (the protostar and disk) \citep[e.g.,][]{whitney2003}. 
The system
has a similar morphology in both its SED and images, to proto-typical Class 0 protostars
like L1527 IRS, B335, and L1157 \citep{tobin2008,stutz2009,looney2007}, all of which are
viewed near edge-on.
However, toward BHR7-MMS there is
no point-like source visible between the outflow cavities as in L1527 IRS, which was due to
scattered light from its edge-on disk \citep{tobin2010b}. Moreover, 
BHR7-MMS exhibits a brightness asymmetry with respect to the north and south cavities.
The dark lane that is apparent from the near-infrared to 8~\micron, the two 
clearly defined scattered light cavities, and the SED shape indicate that the protostar most likely
viewed at an inclination $>$ 60\degr\ but $<$ 90\degr\ \citep{whitney2003b}. It is difficult to more precisely
determine the inclination from the SEDs and images alone due to degeneracies and asymmetry of the envelope 
\citep{furlan2016}. Finally, there are two prominent, apparently symmetric shock
knots located in the center of the outflow cavities. These knots are brightest in 
Ks and 4.5~\micron\ due to the presence of a shock-excited H$_2$ emission line within these bands. The
symmetry of these knots is quite striking, they are both located at a distance of 10\farcs9 ($\sim$4360 AU), indicating a near simultaneous ejection.

The extinction lane in the near-infrared image
appears to have a different position angle with respect to the dark lane
in the IRAC images. This likely reflects some degree of morphological complexity 
in the surrounding envelope \citep[e.g.,][]{tobin2010a}; there is also a hint
 of 8~\micron\ absorption from the surrounding envelope against the mid-infrared background. However,
it is also clear that the system is externally illuminated, likely from a ultraviolet
emitting source because of the limb-brightened 8~\micron\ emission that 
is likely 
from polycyclic aromatic hydrocarbons (PAHs).

\begin{figure}
\begin{center}
\includegraphics[scale=0.45]{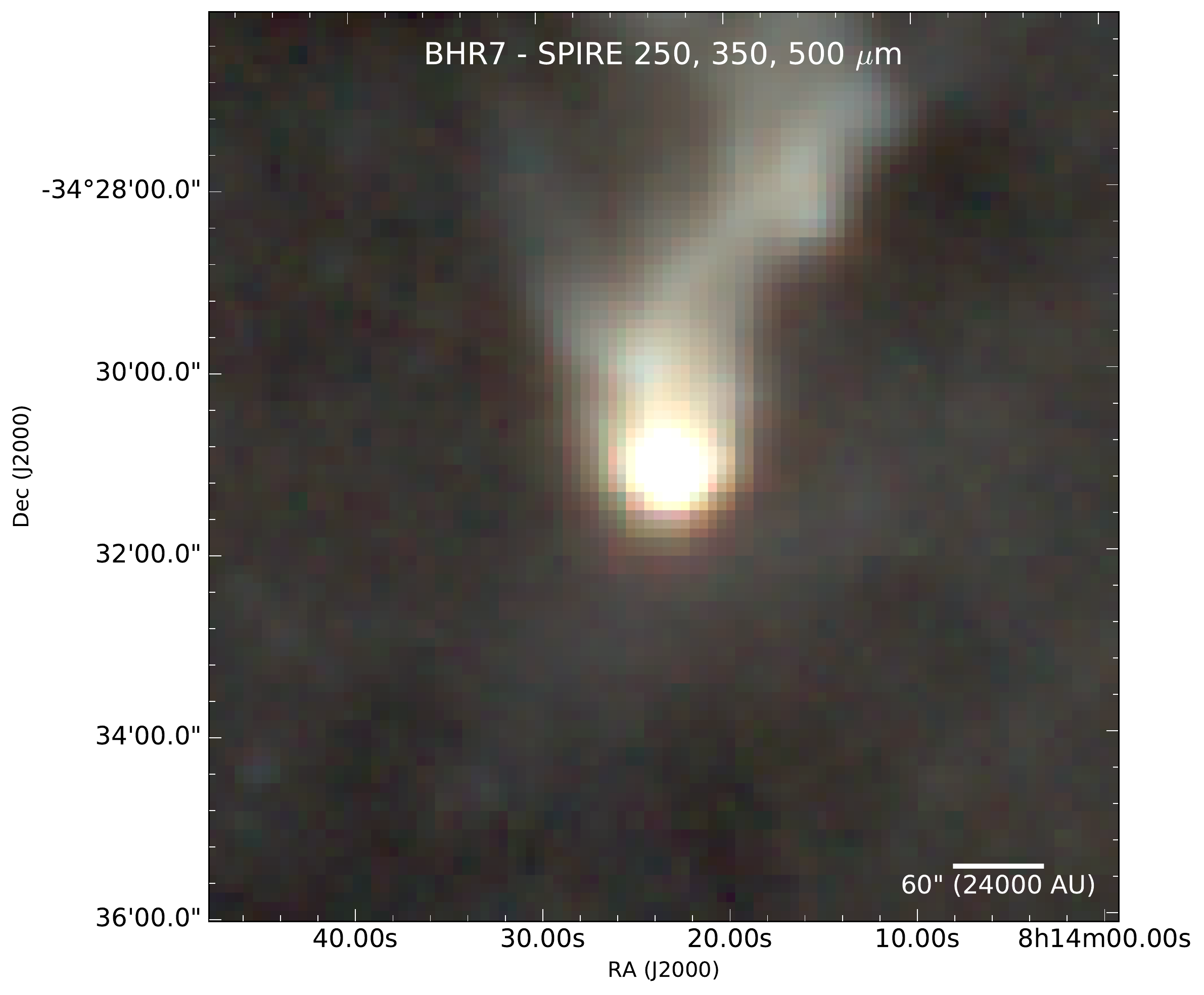}
\includegraphics[scale=0.45]{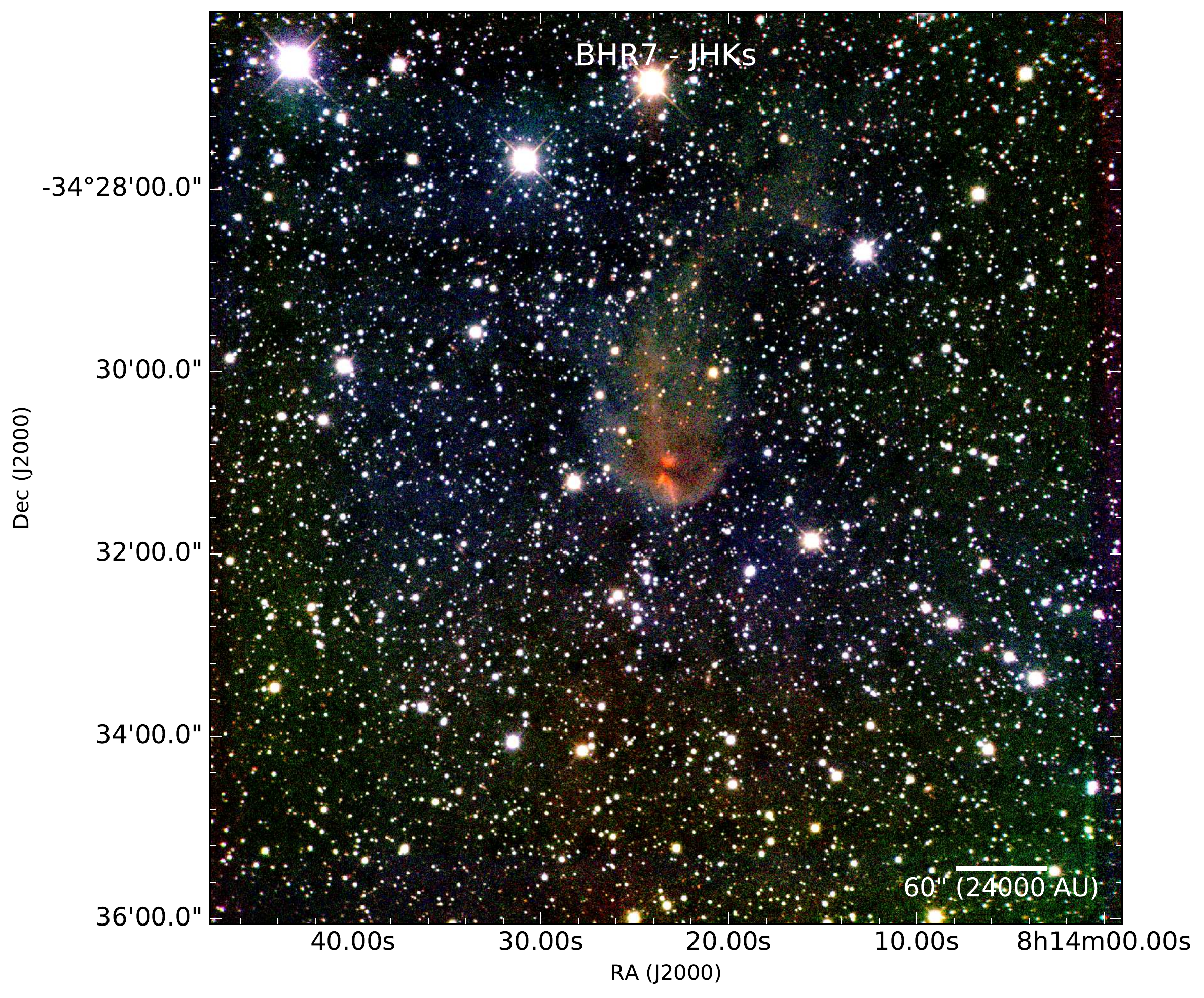}

\end{center}
\caption{\textit{Herschel} SPIRE image at 250, 350, and 500~\micron\ (top) and
a wider-field J, H, and Ks image (bottom); the color ordering of the wavelengths
is blue, green, and red, respectively. The SPIRE image shows the extended cloud and
the dense core at the base where the protostar is forming. Some of the extended
SPIRE emission has corresponding diffuse scattered light in the near-infrared as
shown in the JHKs image.}
\label{bhr7-nir-spire}
\end{figure}

We show the \textit{Herschel} SPIRE data in Figure \ref{bhr7-nir-spire} over a larger field that encompasses
the dense core where the protostar is forming and a lower density tail
extending north. 
We also show a view of the same
field in the near-infrared and there is diffuse scattered light associated with some of the
extended submillimeter emission. This view also shows that the cloud is extended in the direction of the outflow
and that BHR7-MMS has formed at the end of this filamentary structure.

\subsection{Spectral Energy Distribution}

\begin{figure}
\begin{center}
\includegraphics[scale=0.8]{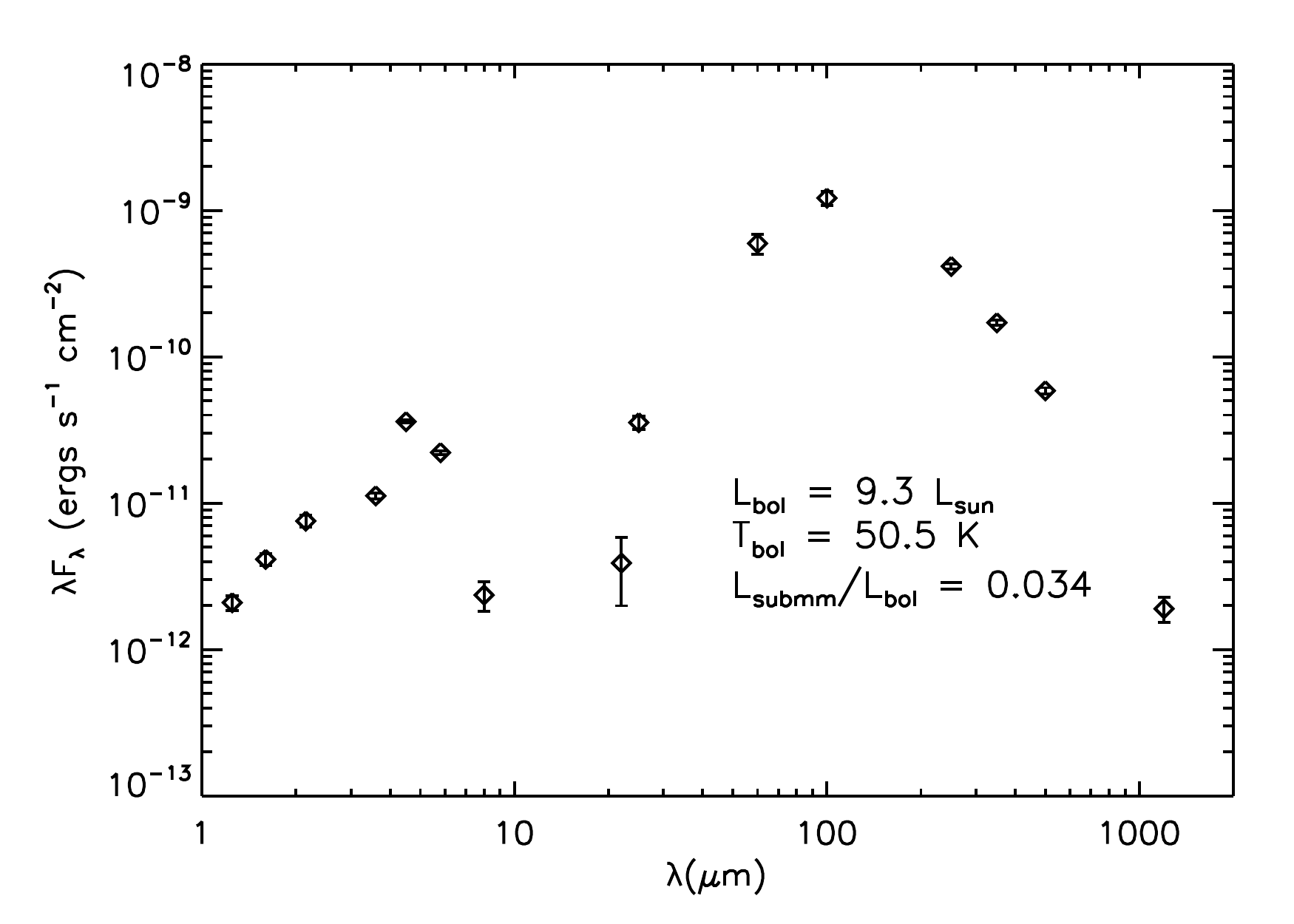}
\end{center}
\caption{Spectral Energy Distribution for BHR7. The 3.6 to 8.0~\micron\ measurements are
from \textit{Spitzer} IRAC, the 25 to 100~\micron\ measurements are from IRAS, the 250 - 500~\micron\ 
measurements are from \textit{Herschel} SPIRE, and the 1.2~mm datapoint is from
SEST SIMBA. From these data, we measured a bolometric luminosity of 9.3~L$_{\sun}$ and a bolometric
temperature of 50.5~K. The ratio of submillimeter luminosity to bolometric
luminosity of 0.034 meets criteria for a Class 0 protostar \citep[$>$0.005; ][]{andre1993}.}
\label{sed}
\end{figure}

Using the near-infrared, \textit{Spitzer} IRAC, IRAS, SPIRE, and SEST SIMBA data, we
constructed the SED of this source in order to determine its evolutionary 
state and plot the SED in Figure \ref{sed}. We
measure a bolometric luminosity of 9.3~L$_{\sun}$, a bolometric temperature 
(T$_{bol}$) of 50.5~K, and a ratio of submillimeter to bolometric luminosity (L$_{bol}$) of 0.034. 
Both of these measurements require integrating the SED for which we use the \textit{tsum} procedure from the
IDL Astronomy library that integrates the SED using trapezoidal integration.
These values indicate that BHR7-MMS is a Class 0 protostar, having both T$_{bol}$ $<$ 70~K and 
L$_{submm}$/L$_{bol}$ $>$ 0.005 \citep{chen1995,andre1993}; both of these metrics are independent
of distance. Furthermore, L$_{bol}$
is significantly above the median for Class 0 protostars, near the top-end of the 
distribution for low-mass systems \citep{dunham2014,fischer2017}.

\subsection{Molecular Line Kinematics}

The SED reveals that this protostar is very young, but molecular lines are key to characterizing
the kinematics of the envelope in order to determine if there is significant rotation or if
it is dominated by infall motion alone.
An isolated Class 0 source with an inclination of $>$60\degr\ is an ideal candidate to examine the
kinematic properties of the infalling envelope. The lack of nearby sources and the clear
geometric orientation simplify the interpretation of molecular line kinematic data.
The observations in Extended and Compact configuration were set-up for the 
mapping of kinematic tracers of the
inner envelope ($^{13}$CO, C$^{18}$O, and H$_2$CO), 
outer-envelope (H$_2$CO and \ntdp), disk ($^{13}$CO, C$^{18}$O), and outflow ($^{12}$CO). 


\subsubsection{The Inner Envelope}

To evaluate the molecular line kinematics, we utilize both moment maps made for the red and 
blue-shifted emission, but also by plotting the position and velocity of the molecular 
line(s). The position-velocity (PV) diagrams of the various molecular line emission are all made with a
PV cut in exactly the east-west direction.
The PV diagram is made by collapsing one spatial 
axis on to the other, summing a strip of specified width, thereby 
transforming the 3D datacube (position-position-velocity) 
into a 2D image (position-velocity).
While this position angle of the PV cut (90\degr\ east
of north) is not exactly orthogonal to the outflow (off by 5.9\degr, see Section 3.8), the PV cut direction
matches the blue and red-shifted emission peaks of the various molecules.

\begin{figure}
\begin{center}
\includegraphics[scale=0.5]{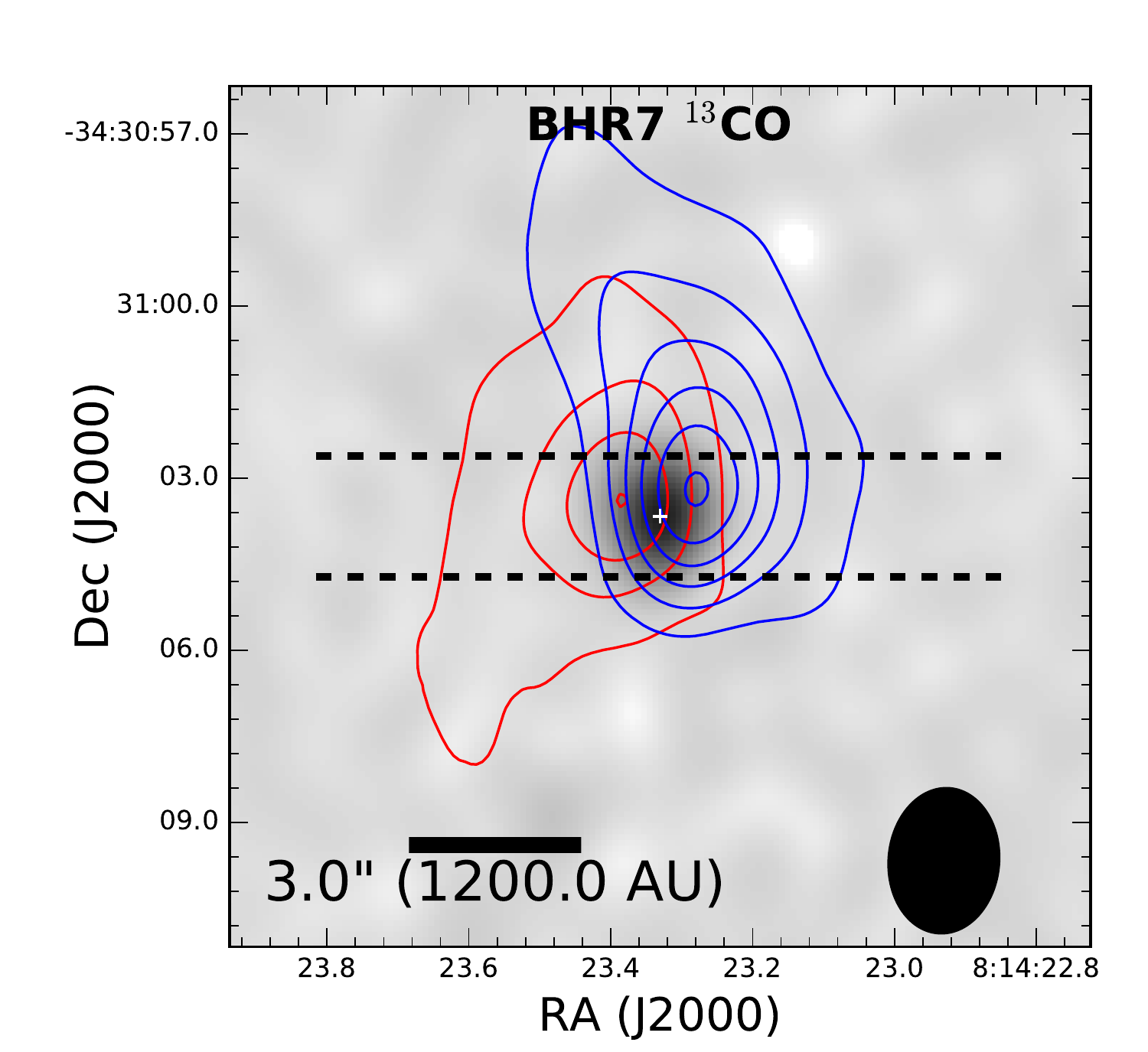}
\includegraphics[scale=0.5]{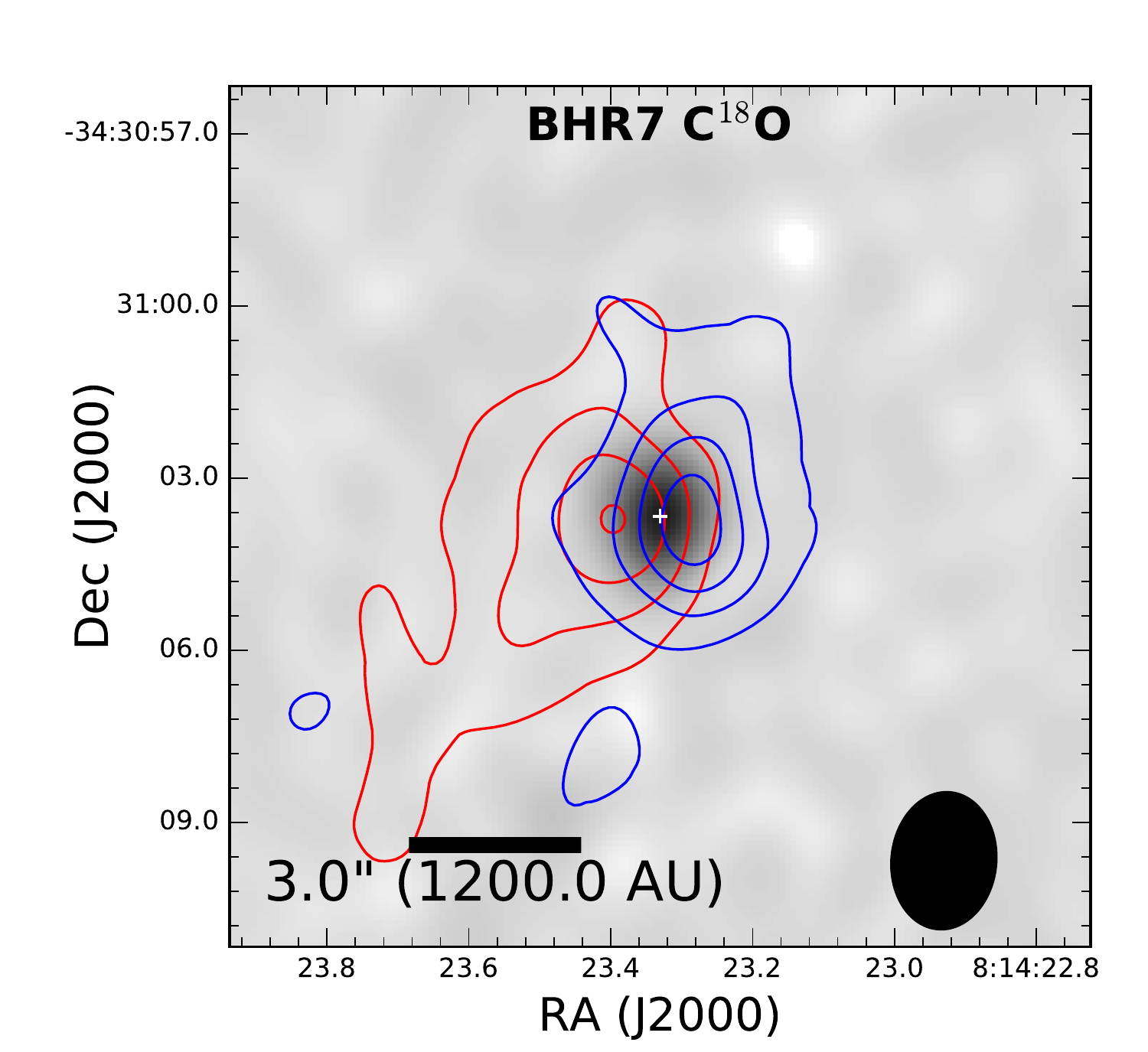}
\includegraphics[scale=0.4]{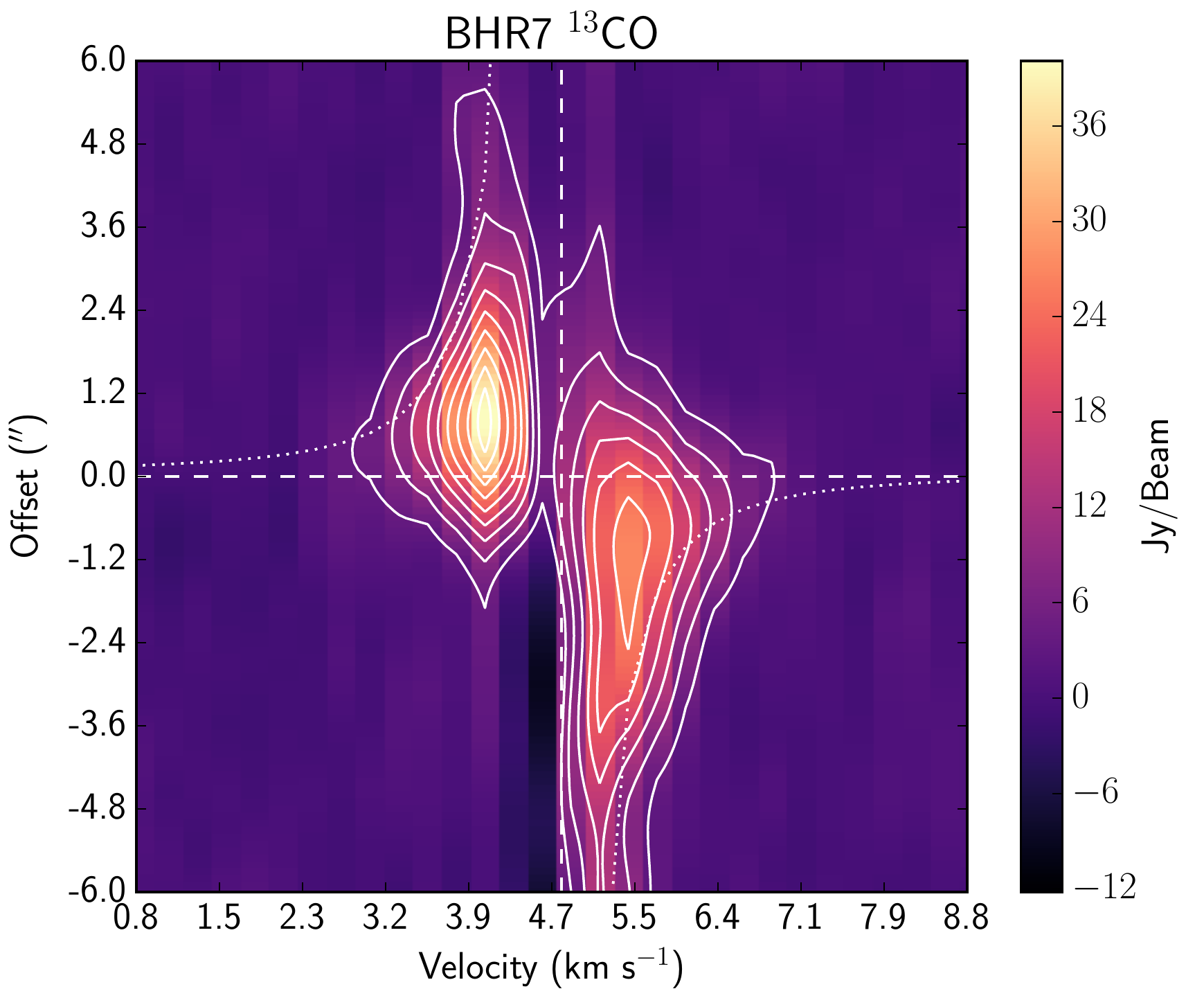}
\includegraphics[scale=0.4]{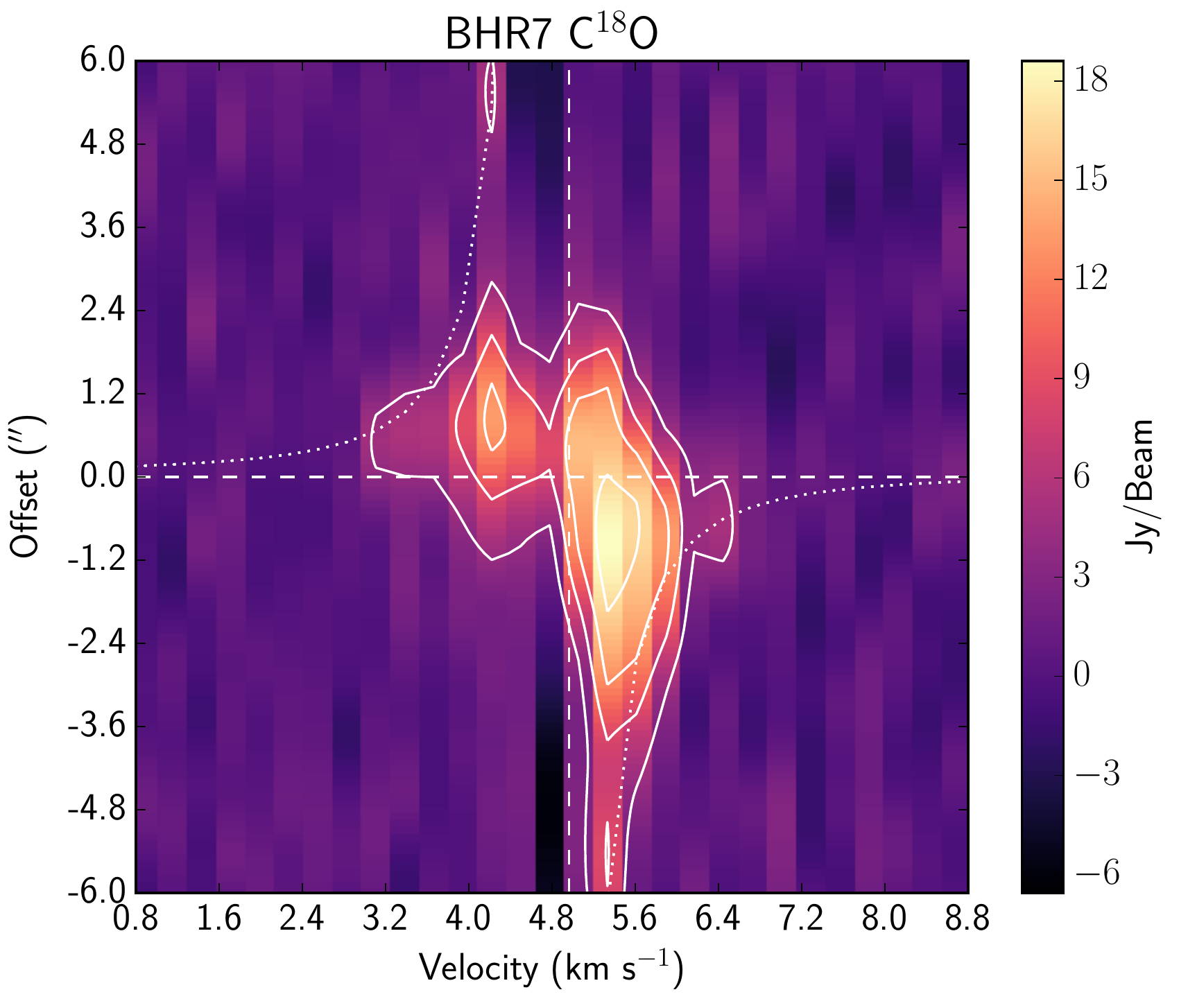}
\end{center}
\caption{BHR7 $^{13}$CO and C$^{18}$O blue and red-shifted integrated intensity maps (top panels)
for the combination of SMA Extended and Compact observations, overlaid on the continuum
imaging from Extended configuration (grayscale). The two CO isotopologues both clearly
show a velocity gradient in the east-west direction, nearly orthogonal to the outflow PA. The dashed
lines overlaid on the $^{13}$CO integrated intensity map denotes the region used for the PV 
diagram extraction for both $^{13}$CO and C$^{18}$O  (12\arcsec\ length, 2\farcs1 width). 
The PV diagrams are shown for the $^{13}$CO and C$^{18}$O (bottom panels), with a Keplerian rotation curve drawn for
a 1.0~M$_{\sun}$ star. The dashed lines mark the source position and system velocity.
The beams for the $^{13}$CO and C$^{18}$O observations are 2\farcs53$\times$1\farcs92 
and 2\farcs39$\times$1\farcs82, respectively. For the $^{13}$CO, the red contours start at 5.0$\sigma$
and increase by 7.5$\sigma$ and the blue contours start at 10.0$\sigma$
and increase by 10.0$\sigma$, where $\sigma_{red}$=0.19~K~\kms\ and $\sigma_{blue}$=0.18~K~\kms. Then
for the C$^{18}$O, the contours start at and increase on 5$\sigma$ intervals; 
$\sigma_{red}$=0.18~K~\kms\ and $\sigma_{blue}$=0.2~K~\kms.
}
\label{co-lowres}
\end{figure}

We show the moment zero (integrated intensity) maps for \thco\ and \cateo\ from the combined Compact and Extended configuration data in Figure \ref{co-lowres}. The integrated 
intensity maps for the blue and red-shifted \thco\ and \cateo\ emission 
clearly show a position shift between the blue and the red on scales $<$1300~AU. The PV 
diagrams shown in Figure \ref{co-lowres} offer more detail on the kinematic structure
of the emission. The PV diagrams show clearly separated blue- and red-shifted components for 
both \thco\ and \cateo, and both molecular lines are resolved out around the source 
velocity. The lower-velocity emission is observed to extend to larger radii
and at a given offset there is a trend for either blue or red-shifted emission to dominate; this
is a clear indication of envelope rotation. There is sometimes 
both blue and red-shifted emission at a given position. The super
position of both blue and red-shifted emission at the same position is indicative of radial
infall motion in an axi-symmetric system \citep{tobin2012a, sakai2014}. We also notice that toward the
source position there is higher velocity emission evident in both \thco\ and \cateo. Given that
there is an indication of rotation at larger radii, this feature likely reflects the increased
rotation velocity (or spin-up) of infalling material due to conservation of angular momentum.

\begin{figure}
\begin{center}
\includegraphics[scale=0.5]{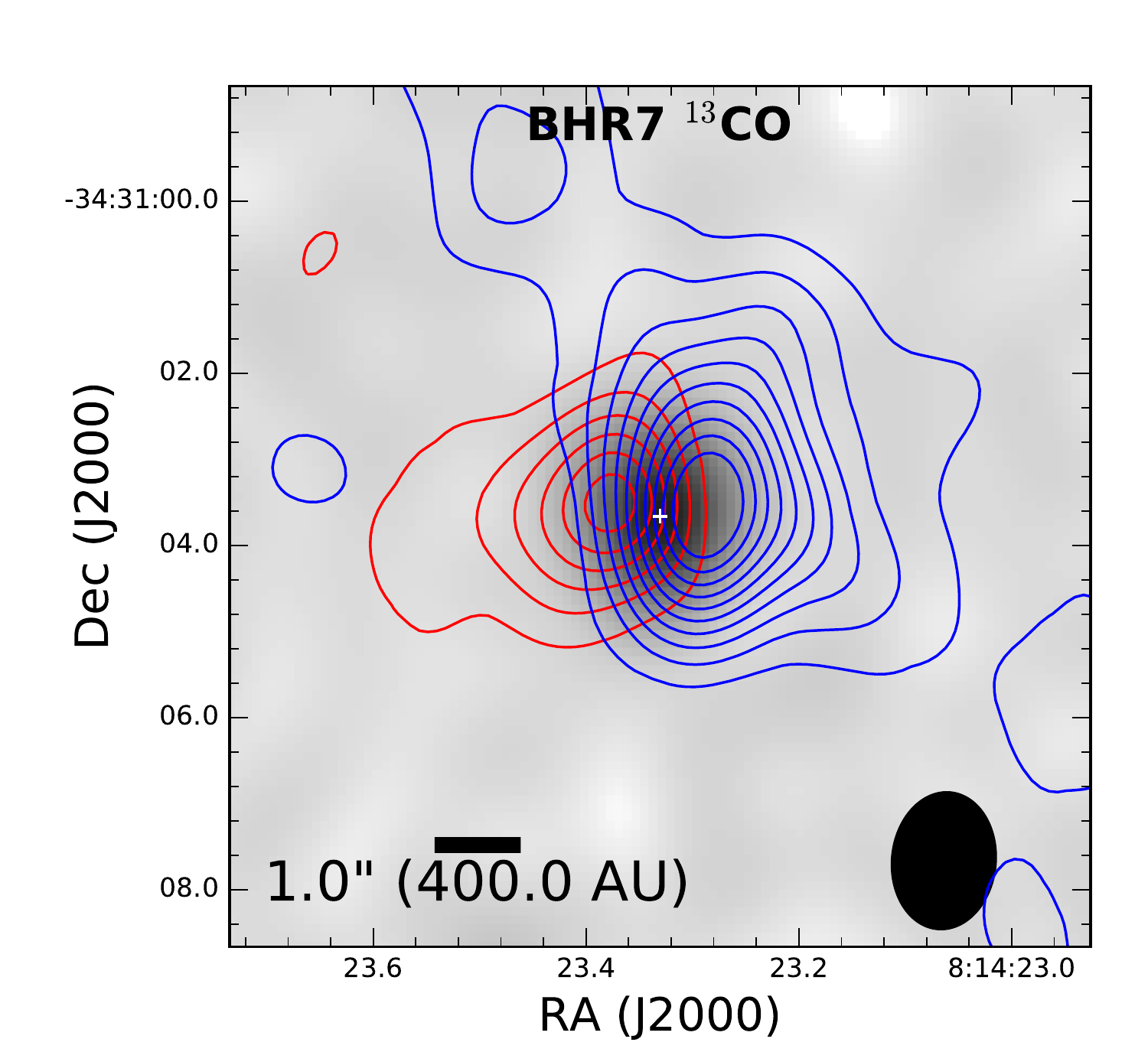}
\includegraphics[scale=0.5]{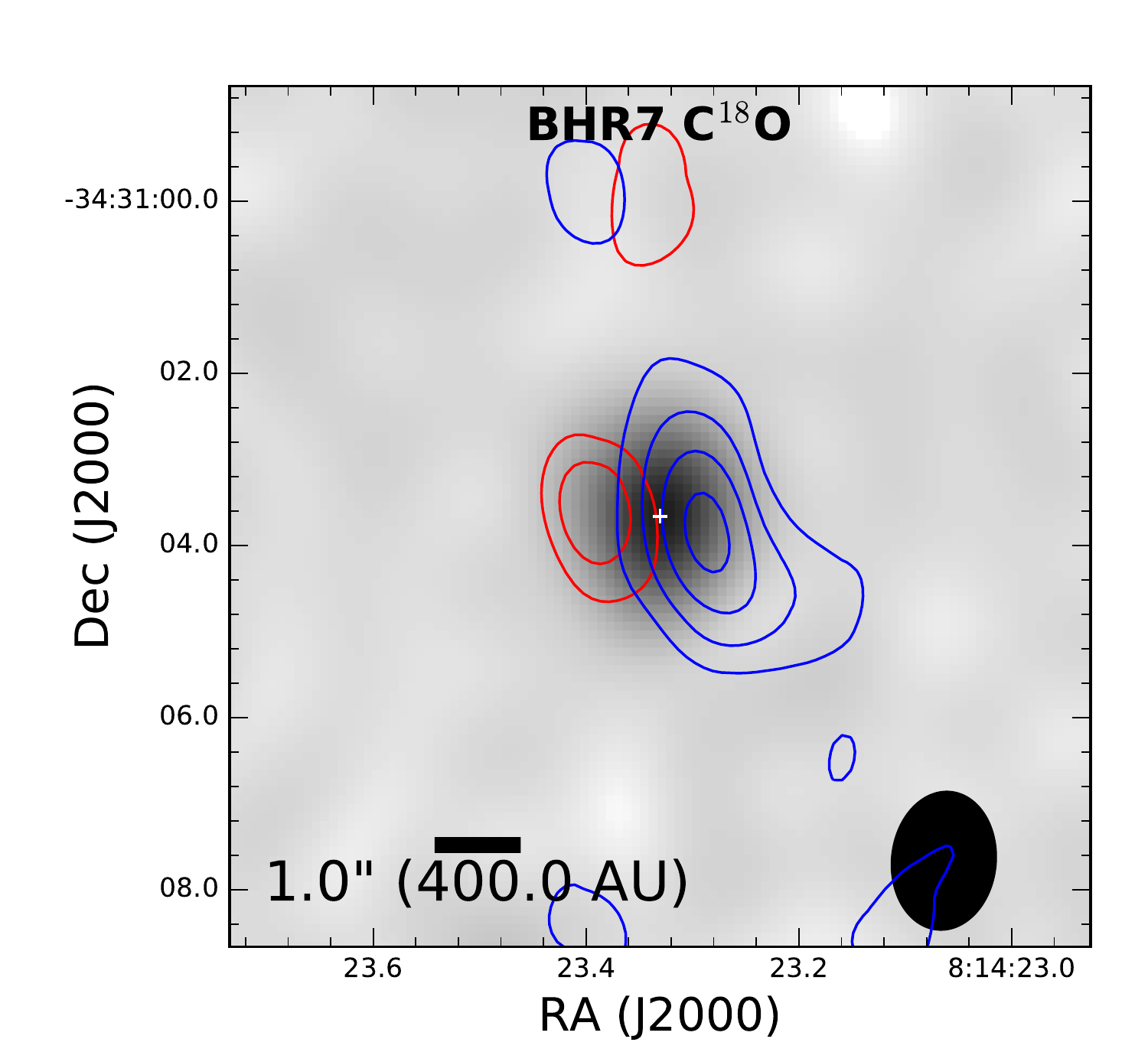}
\includegraphics[scale=0.4]{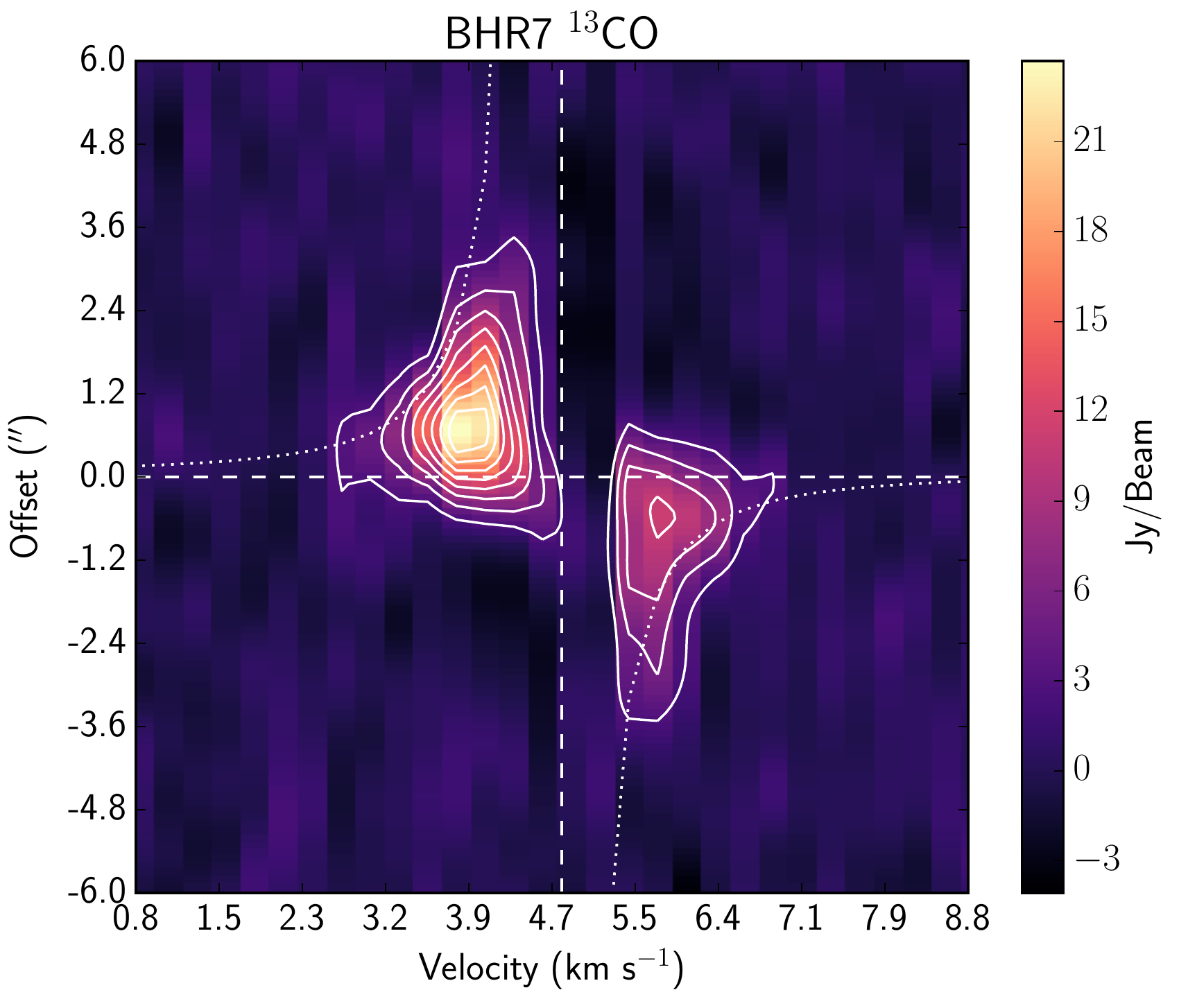}
\includegraphics[scale=0.4]{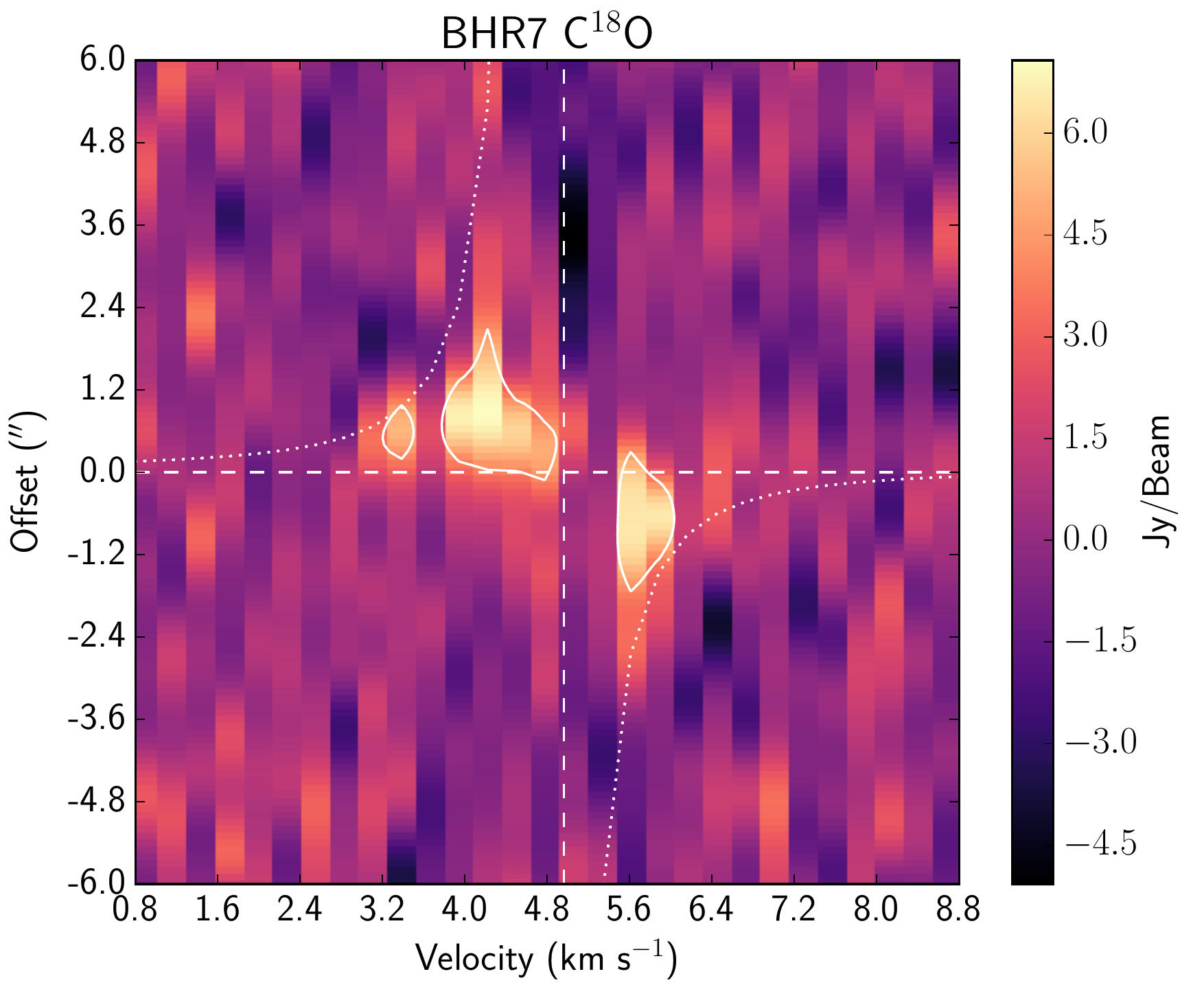}

\end{center}
\caption{BHR7 $^{13}$CO and C$^{18}$O blue and red-shifted integrated intensity maps
from the SMA Extended configuration only (upper panels), overlaid on the continuum
imaging from extended configuration (grayscale). Like the lower resolution data,
both $^{13}$CO and C$^{18}$O trace a smaller-scale velocity gradient orthogonal
to the outflow direction, centered on the continuum position.
The PV diagrams shown in the bottom panels for the $^{13}$CO and C$^{18}$O are extracted using
the same region in Figure \ref{co-lowres}, with a Keplerian curve for
a 1.0~M$_{\sun}$ star drawn (dotted line). The dashed lines mark the source position and system velocity.
The beams for the $^{13}$CO and C$^{18}$O observations are both 1\farcs6$\times$1\farcs2.
For the $^{13}$CO, the red and blue contours start at and increase by 3.0$\sigma$, 
where $\sigma_{red}$=0.43~K~\kms\ and $\sigma_{blue}$=0.51~K~\kms. Then
for the C$^{18}$O, the red contours start at 3$\sigma$ and increase by 2$\sigma$, and
the red contours start at 3$\sigma$ and increase by 1$\sigma$;
$\sigma_{red}$=0.52~K~\kms\ and $\sigma_{blue}$=0.63~K~\kms.
}
\label{co-highres}
\end{figure}

To examine the increased velocity or spin-up observed at smaller radii, we make the 
same blue and red-shifted integrated intensity maps from the Extended configuration data only,
see Figure \ref{co-highres}. The same general features are observed for the high-resolution
data as the low-resolution, but on smaller scales. The higher resolution results in more
large-scale emission being filtered-out and the compact emission toward the source is 
more prominent. In the PV diagrams, the \thco\ emission prominently shows the increased
velocity at small radii. The \cateo\ shows these features as well, but with lower
S/N. These data will be analyzed in more detail in 
Section 3.4.

\begin{figure}
\begin{center}
\includegraphics[scale=0.5]{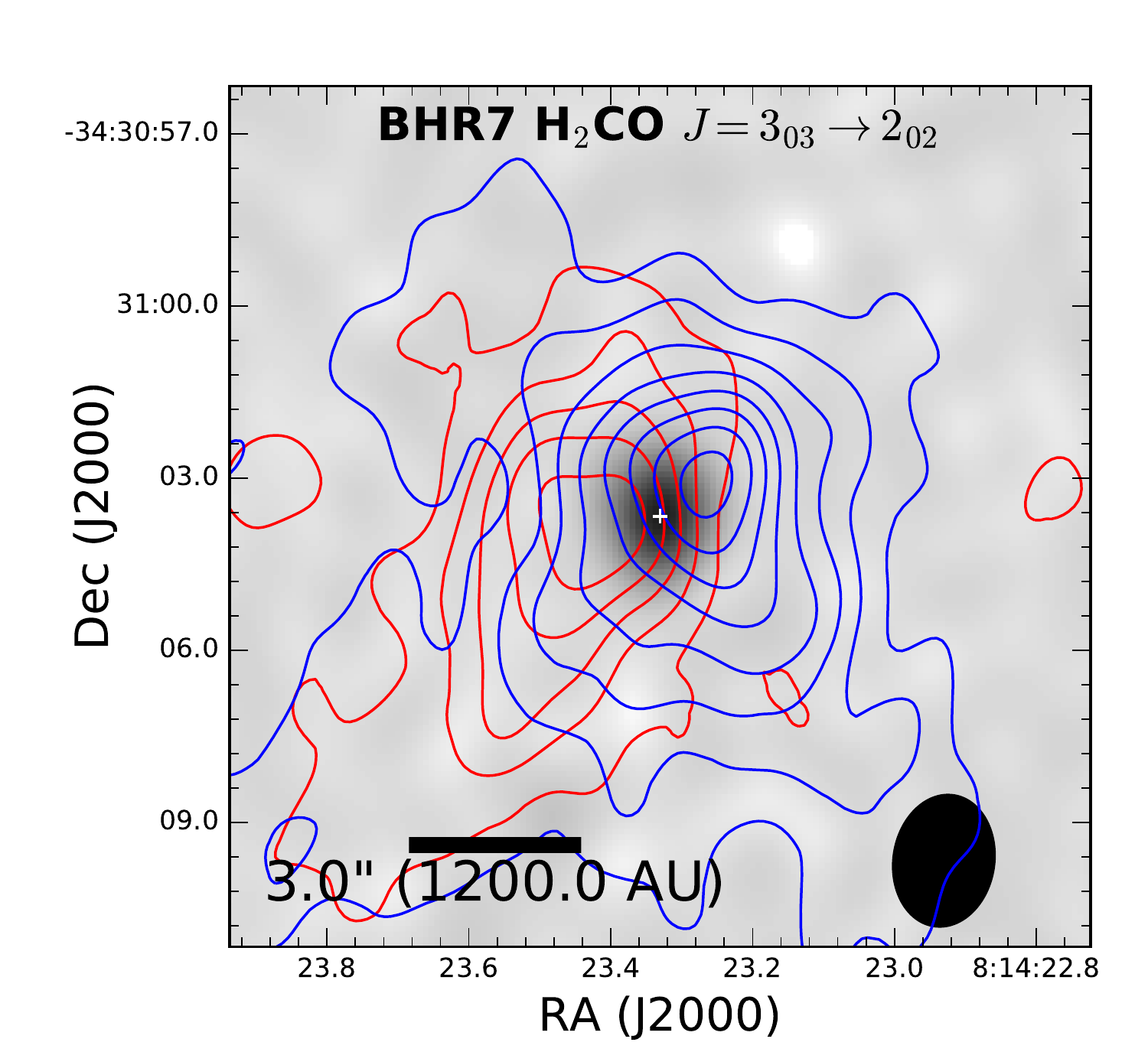}
\includegraphics[scale=0.5]{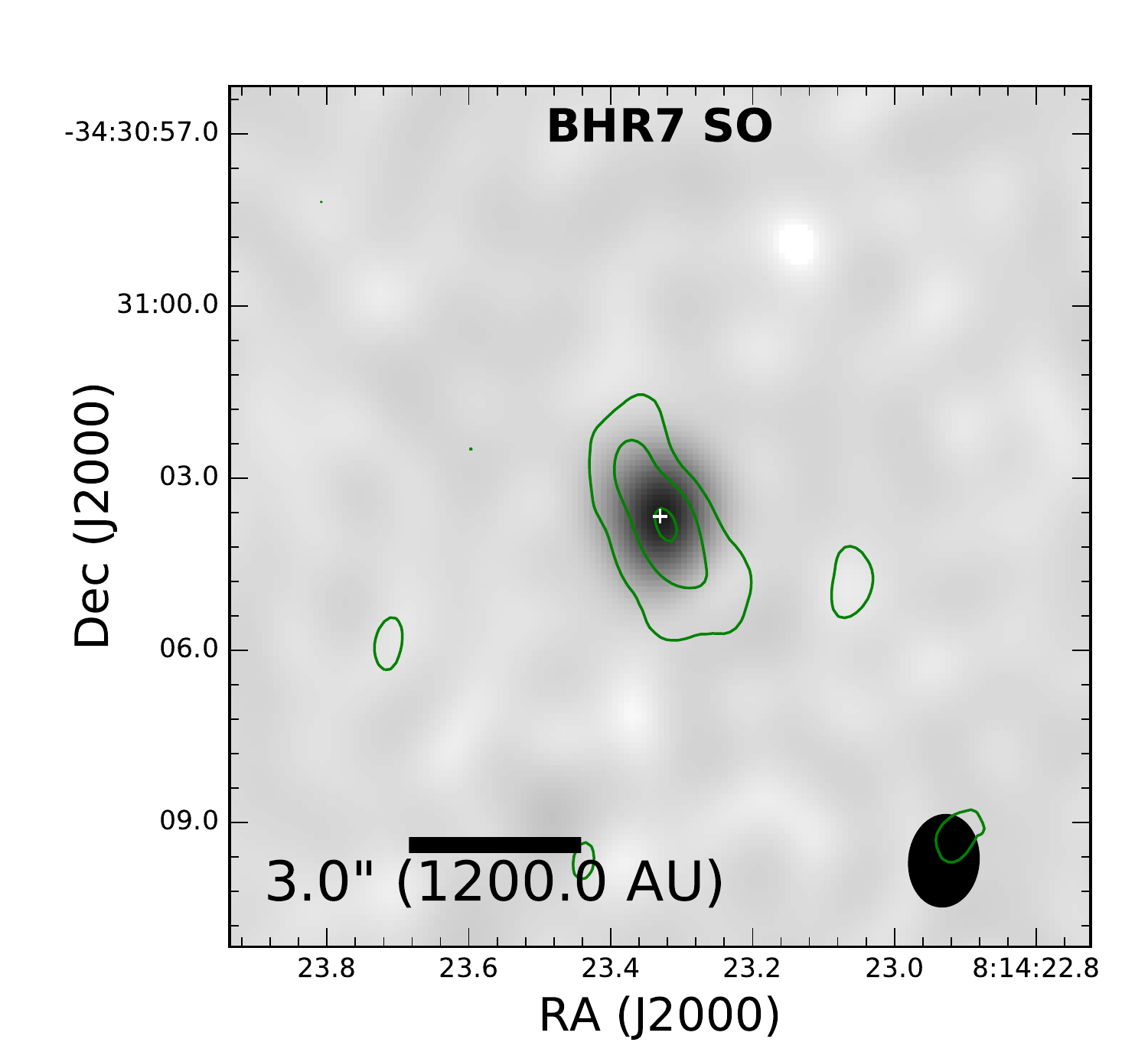}
\includegraphics[scale=0.4]{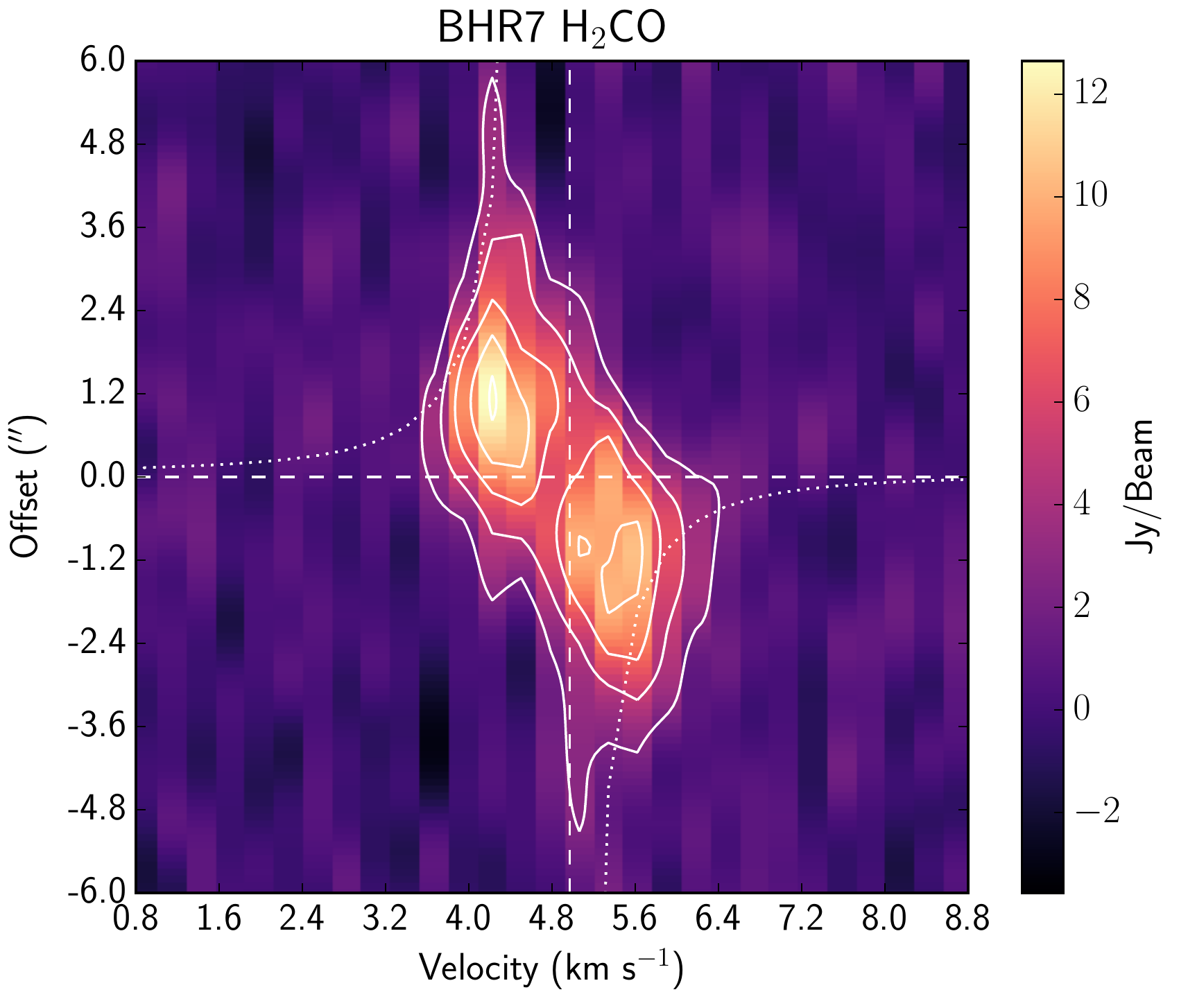}
\includegraphics[scale=0.4]{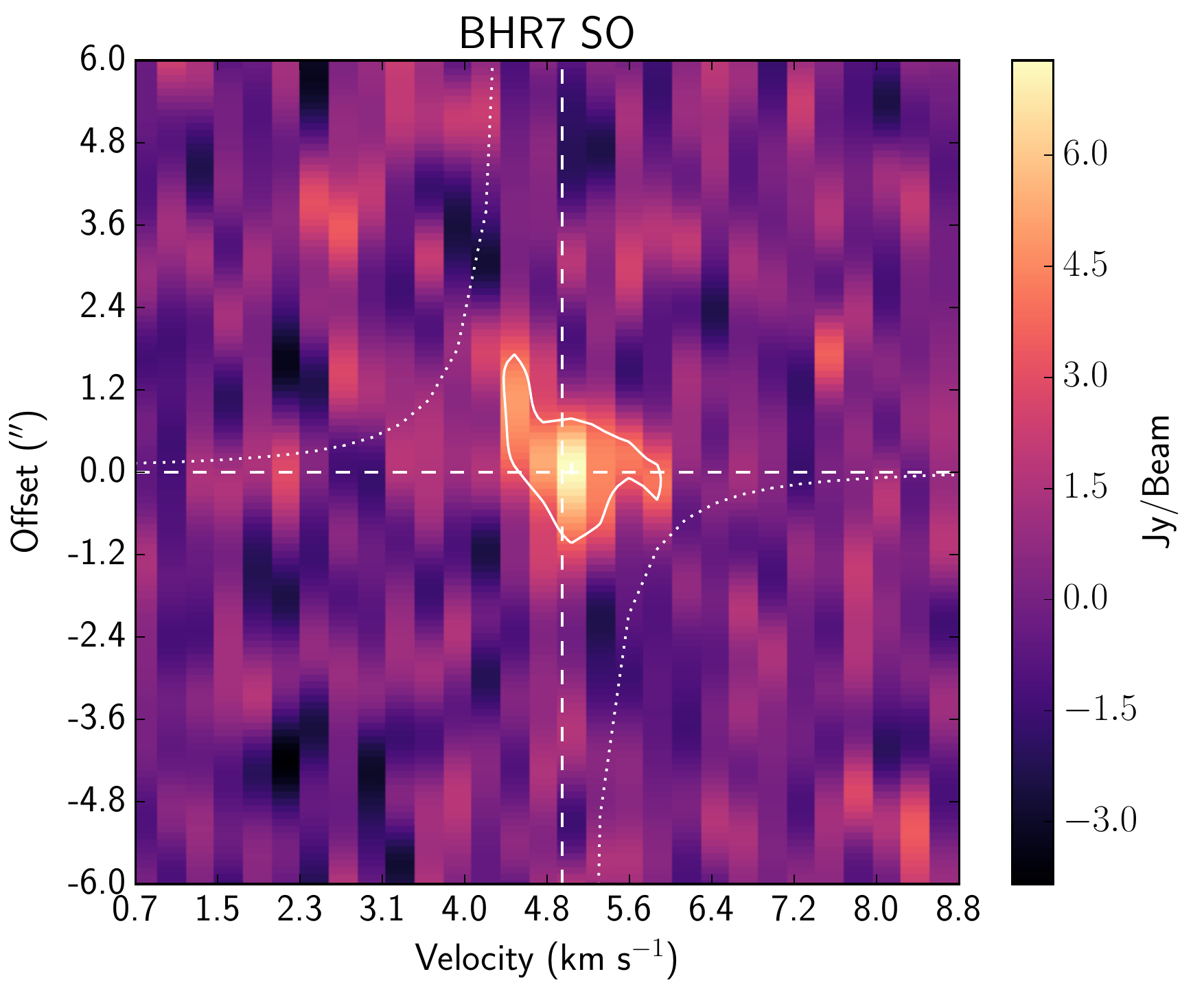}
\end{center}
\caption{SMA H$_2$CO (upper left) and SO (upper right) integrated intensity contours
overlaid on the 1.3~mm continuum (grayscale).
The H$_2$CO is a combination of Extended and Compact observations, while the SO is 
Extended configuration only. The H$_2$CO clearly shows a velocity gradient, in the same
sense as the CO isotopologues, but with an apparent position angle that is slightly larger.
The SO emission is compact and a clear blue and red-shifted separation is not evident at this
resolution and sensitivity. 
The PV diagrams shown for H$_2$CO and SO (bottom panels) are extracted using
the same region shown in Figure \ref{co-lowres} with a Keplerian curve for
a 1.0~M$_{\sun}$ star drawn (dotted line). The H$_2$CO does not strongly trace 
the higher velocities of the inner envelope, but H$_2$CO does trace
the infalling/rotating envelope beyond the extent of the \thco\ and \cateo\ emission very well. The SO appears compact in velocity space.
The dashed lines mark the source position and system velocity.
The H$_2$CO beam is 2\farcs3~$\times$~1\farcs75 and the SO beam is 1\farcs6$\times$1\farcs2.
The SO contours start and increase on 3$\sigma$ intervals where $\sigma$=0.7~K~\kms.
For the H$_2$CO, the contours start and increase on 3$\sigma$ intervals and $\sigma_{red,blue}$=0.18~K~\kms.
}
\label{h2co-so}
\end{figure}

In addition to the CO isotopologues, we also examined the H$_2$CO and SO molecular lines
for indications of kinematic structure on the scale of the inner envelope and disk. \citet{sakai2014}
detected SO toward L1527~IRS that could be tracing an accretion shock
in the outer disk and H$_2$CO traced
both the inner infalling envelope and disk. 
The H$_2$CO emission shows a clear velocity gradient
from the blue and red-shifted integrated intensity maps shown in Figure \ref{h2co-so}; however, the
PV diagram does not show the same high-velocity features toward the source position observed 
in the CO isotopologues (Figure \ref{co-highres}). The PV diagram shows a relatively linear change in velocity
with position in the PV diagram, with some superposition of 
both red and blue shifted emission at a given
position. Furthermore, less emission is resolved-out at line center indicating that H$_2$CO emission is mainly
coming from the inner envelope and there is not significant large-scale H$_2$CO emission in the surrounding
globule. 

SO emission is detected at small radii and near the source velocity as shown in Figure \ref{h2co-so}. 
The integrated intensities are low, such that a separated blue and red-shifted 
integrated intensity map is not practical to generate.
The brightest SO emission being near line-center is consistent with the observations of 
L1527; however, there is a slight extension (beyond what is expected from the beam) along
the direction of the outflow rather than orthogonal to it. While we cannot completely
rule-out an outflow shock component, an outflow shock is not expected to
produce emission at the source velocity without a higher velocity component.
Thus, we cannot firmly associate the SO emission of BHR7 to originate from small-scales like L1527
at the resolution and sensitivity of the data in hand.

\subsubsection{Outer Envelope}

\begin{figure}
\begin{center}
\includegraphics[scale=0.3]{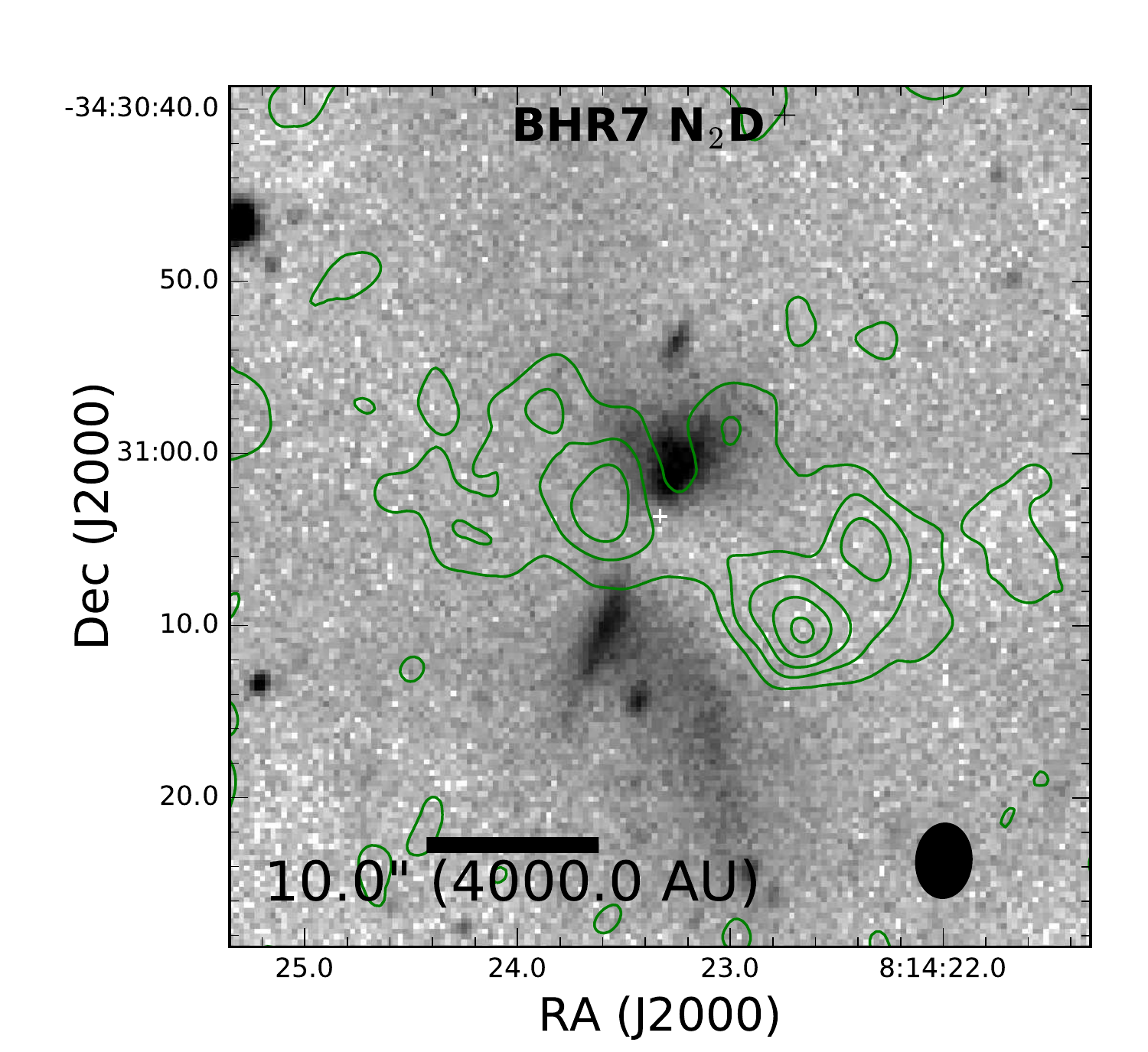}
\includegraphics[scale=0.3]{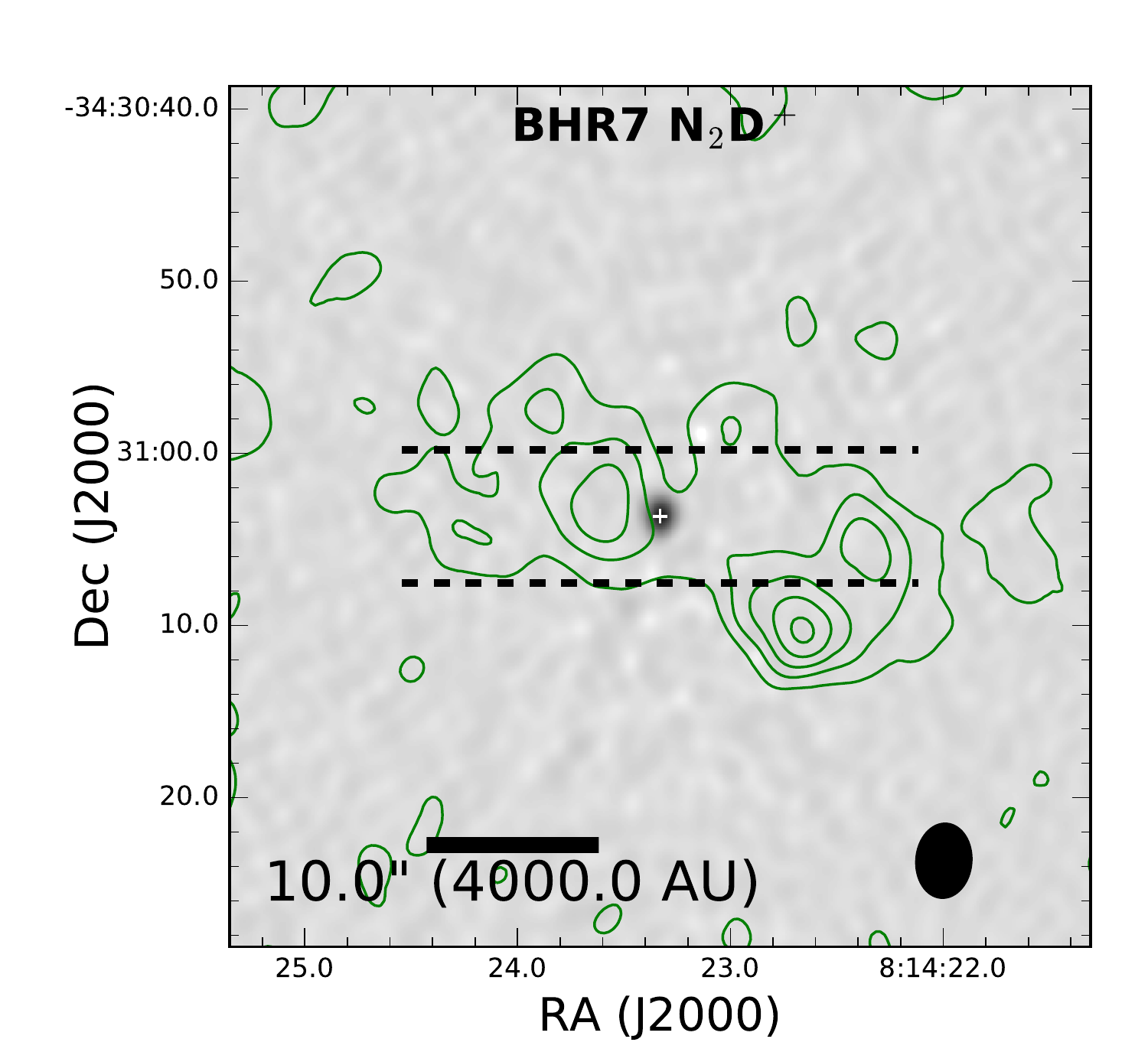}
\includegraphics[scale=0.3]{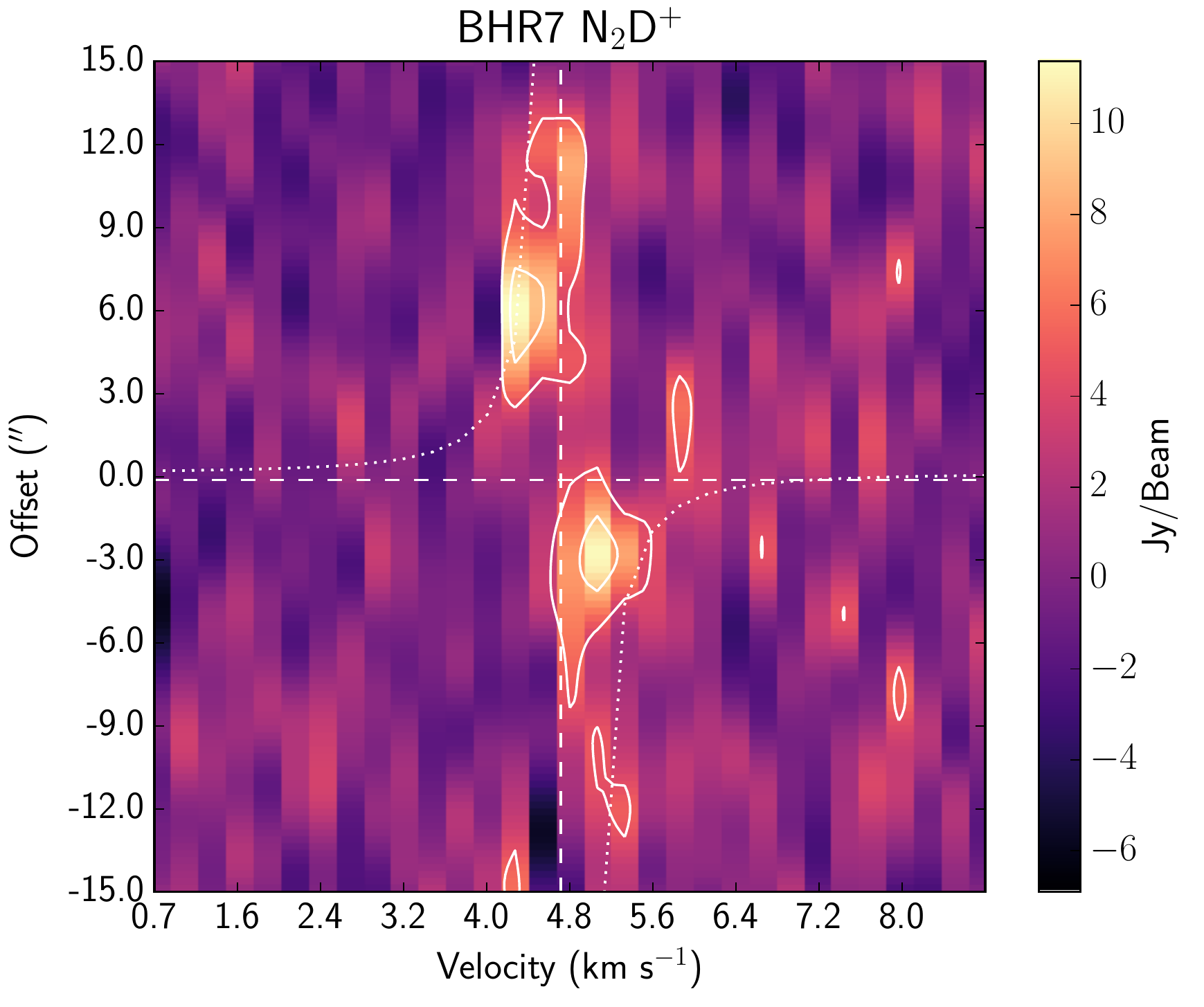}
\end{center}
\caption{SMA \ntdp\ ($J=3\rightarrow2$) observations from combination of 
Extended and Compact observations; however, the data are tapered to the resolution
of the compact configuration to increase the sensitivity to large structures.
The left panel shows the \ntdp\ integrated intensity contours overlaid on 
the Ks-band image, highlighting the coincidence of the thick dark lane. The
middle panel shows the \ntdp\ integrated intensity contours overlaid on the 1.3~mm
continuum.
The \ntdp\ emission is tracing the cold outer envelope where CO is frozen-out
and formation of H$_2$D$^+$ and \ntdp\ is efficient. The peaks of the \ntdp\ emission 
are off the continuum source, as expected
for a protostellar system with significant internal heating. 
The PV diagram for \ntdp, shown in the far right panel, is 
extracted along the same position angle denoted by the dashed lines
in the middle panel (20\arcsec\ length, 7\farcs75 width). 
The PV diagram also shows a large-scale velocity gradient in the 
envelope surrounding BHR7, possibly tracing rotation on scales outside 
H$_2$CO and CO isotopologue emission.
The beam is 4\farcs3~$\times$~3\farcs2 and the contours start at and increase on 2$\sigma$ intervals where $\sigma$=0.1~K~\kms.
}
\label{n2dp}
\end{figure}

In addition to the probes of the inner envelope, we also observed \ntdp\ ($J=3\rightarrow2$),
a tracer of cold dense gas \citep{crapsi2004, emprechtinger2009,tobin2013}. The \ntdp\ transitions,
like the more abundant \nthp\ molecule, have hyperfine structure. However, in the ($J=3\rightarrow2$)
transition, many of the brightest hyperfine lines are at similar frequencies, and we only detect
the main, blended lines well with the SMA and not the weaker, more widely separated hyperfine lines. We show
integrated intensity maps of the \ntdp\ emission in Figure \ref{n2dp}, overlaid on the
1.3~mm continuum and near-infrared images. The emission 
is clearly coming from larger scales than the CO isotopologues, H$_2$CO, and SO emission.
The PV diagram
shows that there is a velocity gradient in the outer envelope that connects
with the inner envelope velocities probed by the CO isotopologues and H$_2$CO. The velocity gradient
could reflect rotation on this scale, but it has also been proposed that large-scale
velocity gradients could also reflect infall of asymmetric envelope structures \citep{tobin2011,tobin2012a}.

The \ntdp\ appears to follow the direction of the dark lane observed in the near-infrared
and it has a deficit of emission toward the protostar position. The deficit is expected
because \ntdp\ (and \nthp) is destroyed by efficient reactions with gas-phase CO \citep{caselli1999, bergin2001, jorgensen2004}. The
presence of gas-phase CO toward the inner regions of the envelope around BHR7-MMS is 
clear from Figures \ref{co-lowres} \& \ref{co-highres}. Moreover, temperatures around and above the
sublimation point of CO will also inhibit the formation of \ntdp\ by reversing the reaction that 
forms H$_2$D$^+$, the key molecule for low-temperature deuterium chemistry \citep[e.g.,][]{langer1985}. The observed morphology
of \ntdp\ resembles other protostars with resolved observations of this molecule \citep[e.g., L1157;][]{tobin2013}, but the \ntdp\ emission peaks on either side of the protostar do not 
appear to be as symmetric as in L1157. The asymmetry could result from an asymmetric
distribution of envelope material around the protostar.

\subsection{Rotation Curve Analysis}

\begin{figure}
\begin{center}
\includegraphics[scale=0.5]{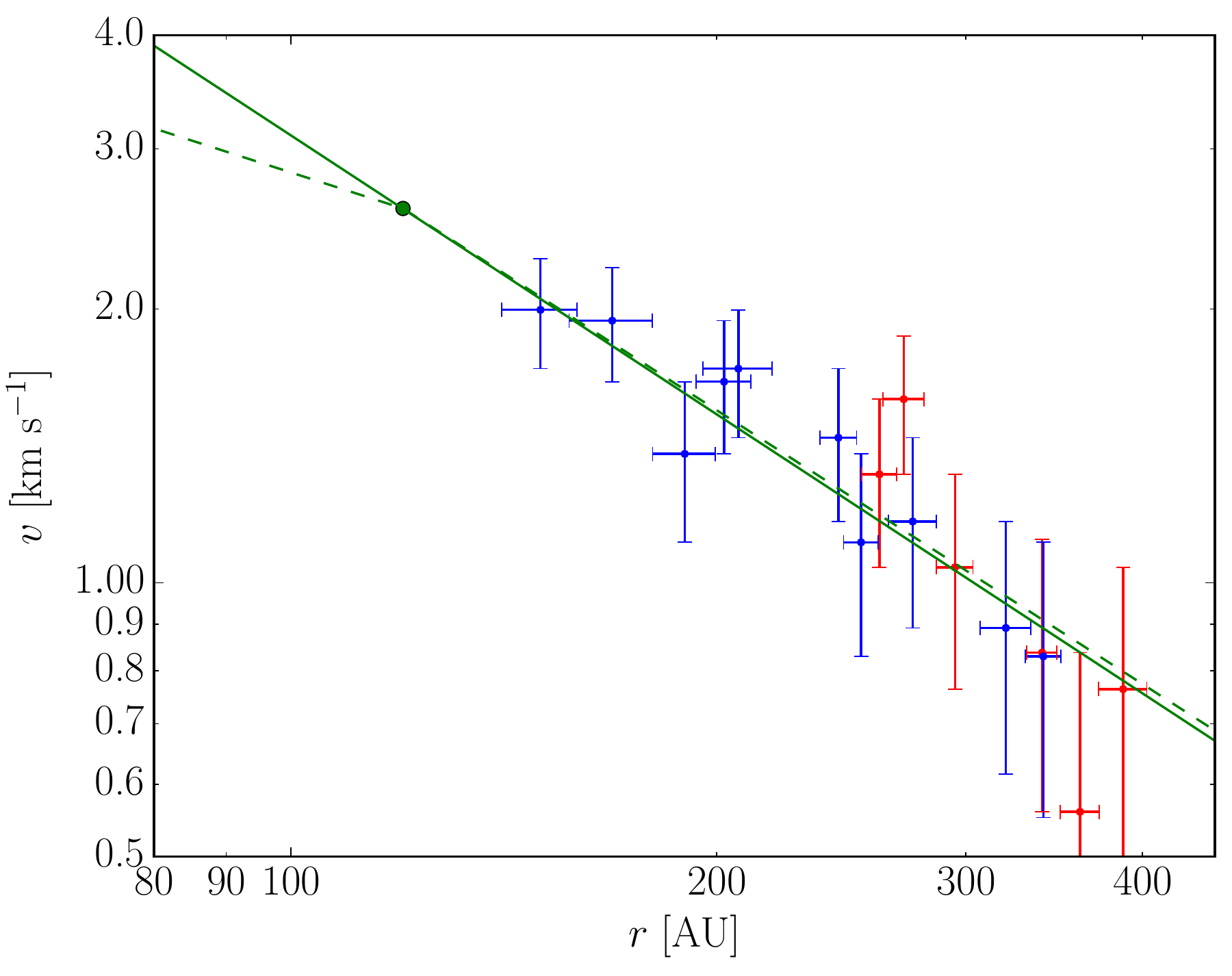}

\end{center}
\caption{The Peak PV diagram of BHR7-MMS including the fitted power law (solid
green line) using $^{13}$CO and C$^{18}$O data. The fit to the power-law is
-1.02$\pm$0.04 and thus consistent with conservation of angular momentum (v $\propto$ r$^{-1}$)
The dashed line shows a broken power law assuming the break point at 120~AU, 2.58~\kms, 
which is denoted by a green dot. With the adopted break point of 120~AU the
inferred mass of the protostar is $\sim$1~$\pm$0.4~M$_{\sun}$. Red data
points are red shifted, blue data points are blue shifted with respect to the source velocity.
}
\label{peakpv}
\end{figure}

While the signature of rotation is evident in the \thco\ and \cateo\ moment maps and PV diagrams, 
further analysis is necessary to determine the radial dependence of the rotation curve. This will enable us to 
determine if the rotation curve reflects conservation of angular momentum (v $\propto$ R$^{-1}$), Keplerian 
rotation (v $\propto$ R$^{-0.5}$), or both on different spatial scales. To examine the rotation curve, we measure the 
position of the \thco\ and \cateo\ intensity peaks in each velocity channel by fitting a 
2D Gaussian. We only select channels with S/N $>$ 4 and where the emission is not extended more than
$\sim$3 resolution elements, enabling the emission peak to be traced as a function of radius.
We have plotted the results with their associated velocities (relative to the source) in log-log space,
yielding a more quantitative means to evaluate the kinematic structure \citep[e.g.][]{tobin2012,yen2013,
ohashi2014}. We refer to this plot as the Peak PV-diagram.
The Peak PV-diagram derived from the high resolution \thco\ and \cateo\ 
data (see Figures \ref{co-lowres} and \ref{co-highres}) is shown in Figure \ref{peakpv}.
As the data points lie on a line in log-log space, we fitted a power-law using a 
Monte Carlo simulation and the least-squares fitting algorithm. The power of the best 
fit is -1.02~$\pm$~0.04, consistent with material rotating with conserved angular momentum
down to a radius of $\sim$170~AU, the smallest radial displacement measured in the molecular line data.
Thus, we cannot identify a Keplerian
rotation signature in the Peak PV-diagram, likely because the Keplerian region is found
on smaller radii than we can examine with our observations. More sensitive observations
will be required to probe the kinematics down to smaller radii to clearly identify a region
of Keplerian rotation.
We discuss the implications of the detected rotation further in Section 4.

\subsection{Sub-arcsecond Resolution SMA Data}

\begin{figure}
\begin{center}
\includegraphics[scale=0.7]{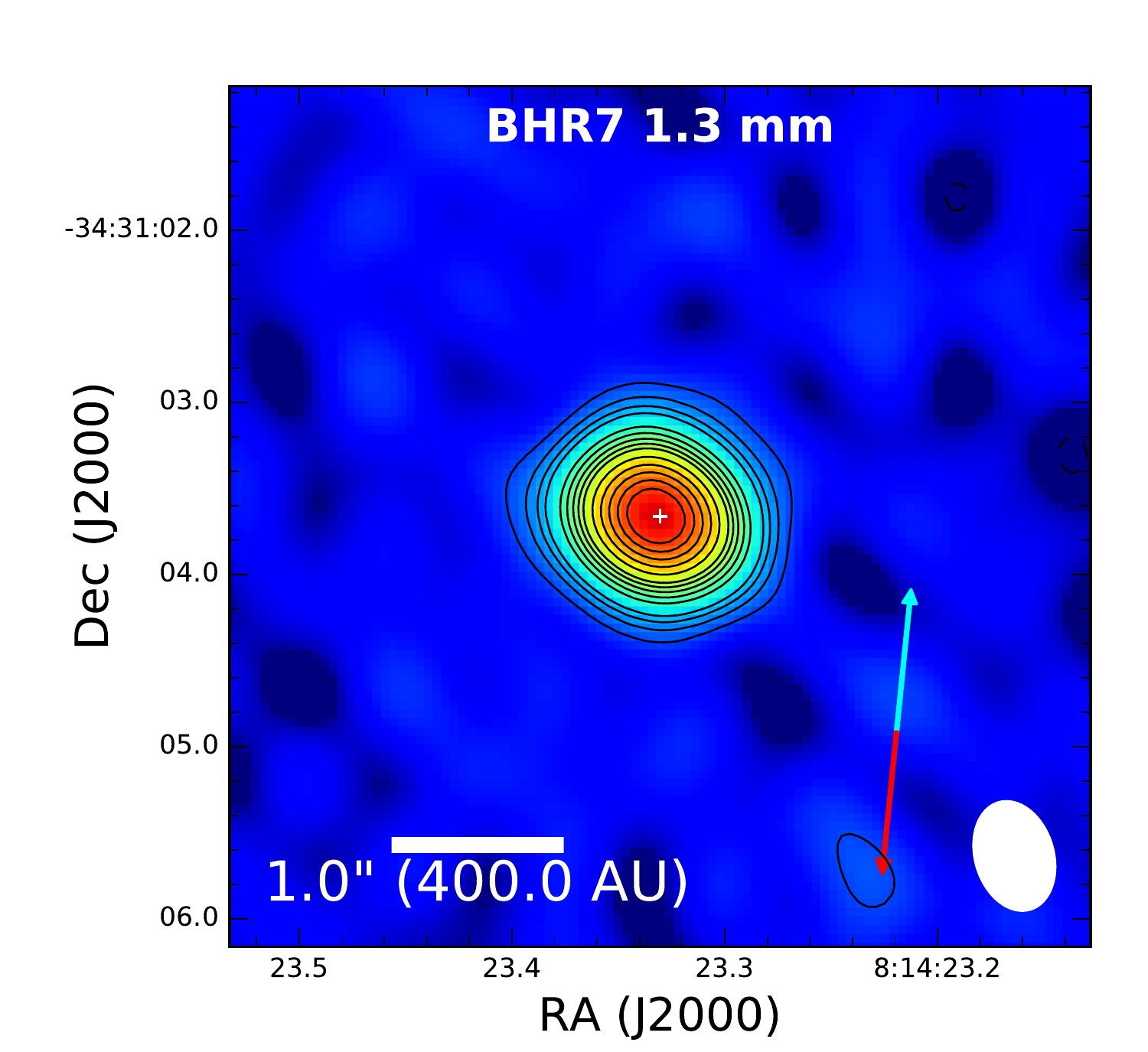}
\end{center}
\caption{SMA 1.3~mm continuum image of BHR7-MMS from Very Extended configuration.
The image shows a 
resolved
dust continuum source embedded within BHR7. The
source is clearly extended orthogonal to the outflow direction, marked by the blue
and red arrows. The width of the continuum source is substantially 
wider than the minor axis of the 0\farcs6~$\times$~0\farcs45 (240~AU~$\times$~180~AU) beam.
The deconvolved size of the continuum source is 0\farcs52$\pm$0.01~$\times$~0\farcs15$\pm$0.04 (208~AU~$\times$~60~AU) with a position angle of 91$\pm$2\degr. This is indicative
of a compact, disk-like structure in BHR7.}
\label{sma-disk}
\end{figure}

We observed BHR7-MMS at the highest resolution available at 1.3~mm 
from the SMA (0\farcs6~$\times$~0\farcs45; 240~AU~$\times$~180~AU) 
to determine if there was evidence of resolved
structure on sub-500 AU scales; 
the inclination of $>$60\degr\
helps simplify the interpretation of the continuum morphology.
The SMA continuum image at the highest
resolution is shown in Figure \ref{sma-disk}; this image is produced using only the 
Very Extended data. The image shows a very strong 1.3~mm detection,
that is extended along the minor axis of the beam, orthogonal to the outflow direction (Section 3.8). 

The deconvolved size of the continuum source is 0\farcs52$\pm$0.01~$\times$~0\farcs15$\pm$0.04 
(208~AU~$\times$~60~AU) with a position angle of 91$\pm$2\degr. This is suggestive
of a compact, disk-like structure around BHR7-MMS with a radius of $\sim$104~AU, at the assumed distance
of 400~pc. If the continuum is tracing a symmetric, geometrically-thin disk, we can estimate the
inclination of the disk by taking the inverse cosine of the minor/major axis obtained from
the Gaussian fit deconvolved from the synthesized beam. Note that this method will break down when the source
inclination approaches 90\degr\ (edge-on) because the disk is not infinitely thin and the inherent
thickness will never result in an inclination that is 90\degr.
With these assumptions and caveats in mind, we estimate the system inclination to
be 73.2\degr, which is consistent with the estimates based on the SED and image morphology.
The integrated flux-density of the source is 197$\pm$2.2~mJy (127$\pm$1~mJy peak);
these measurements also result from the Gaussian fit. 
The lower-resolution Compact data are more sensitive to the extended envelope emission and
we observe an integrated flux density of 273$\pm$49~mJy.

\subsection{Mass Estimates from Dust Continuum}

The mass of the envelope and compact structure in BHR7 can be estimated using the 
flux density of the dust continuum emission observed toward it. 
Under the assumption that
the dust emission is optically thin and isothermal, the dust mass is given by
\begin{equation}
\label{eq:dustm}
M_{dust} = \frac{D^2 F_{\lambda} }{ \kappa_{\lambda}B_{\lambda}(T_{dust}) }.
\end{equation}
D is the distance ($\sim$400~pc), F$_{\lambda}$ is the observed flux density, B$_{\lambda}$ is
the Planck function, T$_{dust}$ is the dust temperature, and $\kappa_{\lambda}$ is the
dust opacity at the observed wavelength. T$_{dust}$ is assumed to be 30~K, consistent
with temperature estimates on $\sim$100~AU scales \citep{whitney2003a} and $\kappa_{\lambda}$
is 0.899, taken from \citet{ossenkopf1994}. We then multiply the resulting value of M$_{dust}$ by
100, assuming the canonical dust to gas mass ratio of 1:100 \citep{bohlin1978}.

The flux density from the Very Extended observations of 197$\pm$2.2~mJy translates to
a mass of 
0.42~M$_{\sun}$ 
and the flux density from the Compact observations of 273$\pm$49~mJy
translates to 
0.51~M$_{\sun}$.
Because the interferometer is missing some flux, we also calculate
the mass from the 1.2~mm SIMBA flux density measurement from SEST. The flux density was 0.76~Jy at
1.2~mm, translating to a mass of 
1.2~M$_{\sun}$, 
assuming the dust opacity spectral index of $\sim$1.8 from \citet{ossenkopf1994}.
If we extrapolate the 1.2~mm flux density to 1.3~mm using the dust opacity 
spectral index of $\sim$1.8, the SIMBA flux density is $\sim$0.55~Jy. This means that a bit less
than half of the continuum emission is coming from sub-arcsecond scales (radii $<$ 200 AU), 
as traced by the Very Extended observations, and the rest is coming from the extended envelope.
Furthermore, we had assumed a temperature of 30~K in the previous calculations, if the average temperature
traced in the SIMBA data was 20~K rather than 30~K, the corresponding mass would be 2.03~M$_{\sun}$. 
The \textit{Herschel} 500~\micron\ also traces cold dust from the envelope and if we assume T=20~K and
the same dust opacity spectral index, the mass is calculated to be 1.55~M$_{\sun}$. The 500~\micron\
data may be more affected by opacity from scales near the disk and this could underestimate the total
mass.

\subsection{Visibility Amplitude Profile}

\begin{figure}
\begin{center}
\includegraphics[scale=0.5]{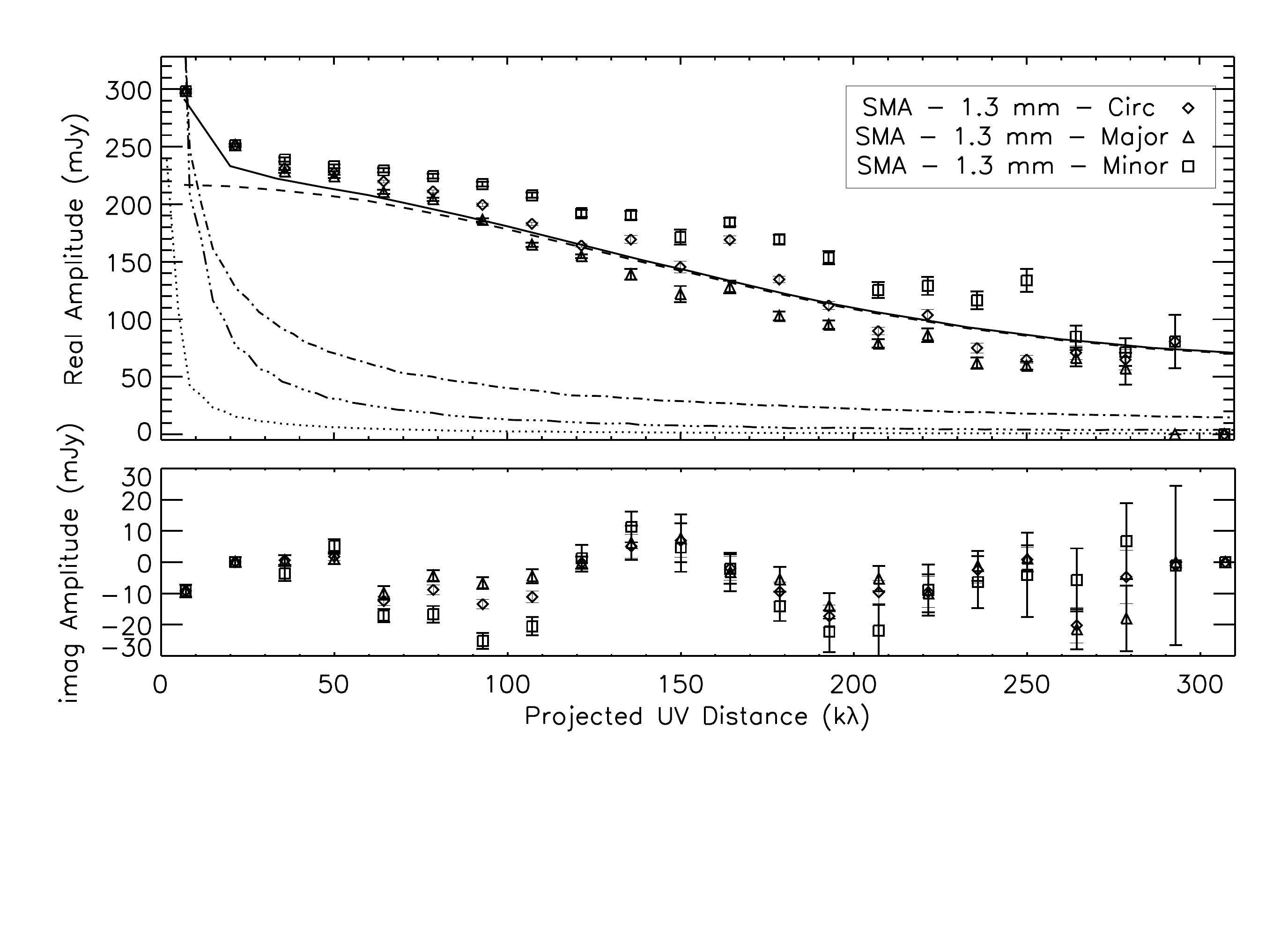}

\end{center}
\caption{Visibility amplitudes from the combined Compact, Extended, and Very Extended datasets.
We show both the circularly averaged visibility amplitudes and cuts 
in the uv-plane 
along the major and minor
axis of the envelope. The major axis decreases more rapidly, demonstrating that the system
is more compact along the direction of the outflow than along the equatorial plane of the disk.
We overlay visibility amplitudes from radiative transfer models of a rotating, 
infalling envelope (dotted line) \citep{ulrich1976,cassen1981} and a 100 AU disk, 
having a surface density profile proportional to R$^{-1}$ (dashed line). The combination
of these two components is shown as the solid line. Assuming that all the emission is
optically thin, the data are consistent with a 1.57 M$_{\sun}$ envelope and a 0.47 M$_{\sun}$ disk.
We also overlay an infalling envelope with a volume density profile proportional to R$^{-2}$ (dot-dashed line) and
a rotationally-flattened envelope (triple-dot-dashed line). Both envelope models do not have an embedded disk included
and they were scaled to match the shortest visibility amplitude.
The imaginary plot on the bottom show some hints of possible non-axisymmetry (systematically non-zero values)
but at low S/N.
}
\label{uvamps}
\end{figure}

The visibility amplitude data themselves can also reveal signatures of structure that are not
readily apparent in the images. We show the visibility amplitude profile toward BHR7-MMS
in Figure \ref{uvamps} for the combined Compact, Extended, and Very Extended datasets.
 We present three different
views of the visibility amplitude data, standard circular averaging, a section of the $uv$-plane 
along the major axis of the source (orthogonal to the outflow) 
and a section along the minor axis of the source (along the outflow).
There is a clear envelope component at short $uv$-distances, where 
the amplitudes steeply rise and then there is a more slowly declining 
component from about 30~k$\lambda$ to 300~k$\lambda$.
 At $uv$-distances $>$50~k$\lambda$ the three points begin to diverge 
and visibility amplitudes along the major axis decline 
more rapidly than those along the minor axis; a similar analysis technique was
also employed by \citep{aso2017}. 
The divergence of the three profiles
is indicative of a flattened
structure. This trend is evident to $uv$-distances $>$200~k$\lambda$ (scales $<$1\arcsec).
 There is also signal at the greatest $uv$-distance we measure, indicating that the structure in BHR7
is not fully resolved. The slowly declining component reflects a more compact 
structure, possibly a disk.

The imaginary visibility amplitudes also show some structure, a symmetric source would have
an imaginary component with zero amplitude. Therefore, we are tracing some asymmetric
structure in the envelope down to scales approaching the disk size. However, it is difficult
to quantify the nature of the asymmetry with our S/N, but we can further analyze the 
real component using simple models for envelope and disk structure. In Figure \ref{uvamps},
we also overlay model visibility amplitude profiles for several 
different density structures:
a disk with a surface density profile of R$^{-1}$ and radius of 100~AU, 
an envelope undergoing rotating collapse \citep{ulrich1976,cassen1981} with a centrifugal radius of 100~AU
(the density profile inside the centrifugal radius is proportional to R$^{-1/2}$ inside of the 
centrifugal radius and proportional to R$^{-3/2}$ outside), a power-law envelope with a 
radial density profile proportional to R$^{-2}$, and the combination of a disk and the rotating
collapse model.

These models are from the 
same set of Hyperion \citep{robitaille2011} models used in \citet{tobin2015}, and we assume that the emission is optically 
thin such that we can freely scale and combine the two components without re-running the 
radiative transfer model.
The two envelope-only models are scaled to 
match the point with the smallest uv-distance, but do not contain an embedded disk. 
This shows that an envelope-only models cannot describe the observational data without a disk component.
Even if the envelope-only models were are scaled up to come closer to matching the small-scale structure,
the shape of visibility amplitude profiles are
inconsistent with the profile observed from baselines longer than 50~k$\lambda$. Thus, the visibility amplitude
profile for a 100~AU disk provides the best match to the data.

The data are consistent with 
an envelope having a centrifugal radius of 100 AU 
and a mass of 
2.1 M$_{\sun}$, within the envelope an
embedded disk with a 100 AU radius and a mass of 0.47 M$_{\sun}$
is needed to reproduce the visibility amplitude profile.
This analysis
demonstrates that the data cannot be fit with an envelope-only model, requiring
a more compact component that is 
consistent with a
a disk. A self-consistent
radiative transfer model of the disk and envelope, 
as well as the SED,
is reserved for future work.

\subsection{Outflow}

\begin{figure}
\begin{center}
\includegraphics[scale=0.4]{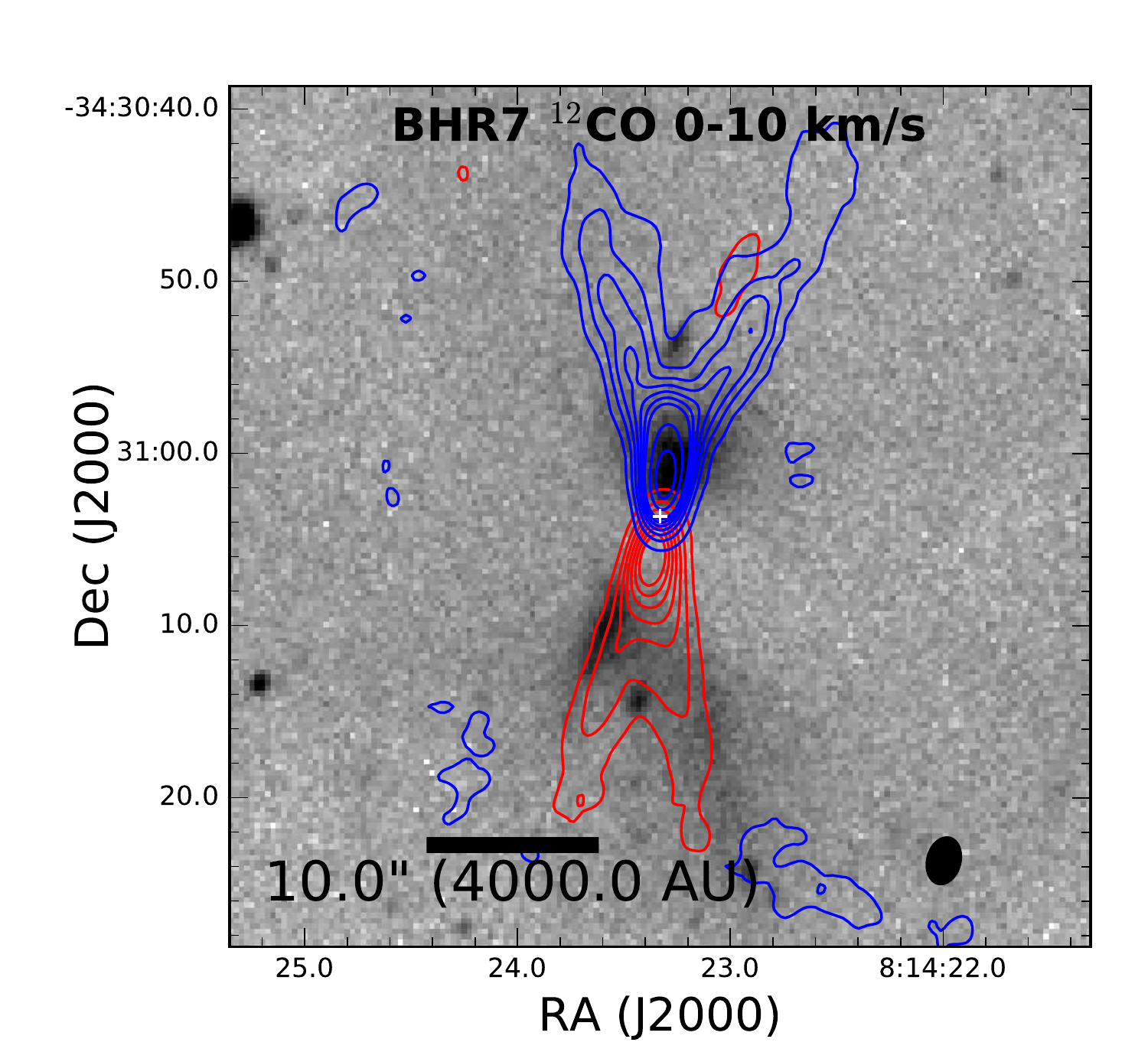}
\includegraphics[scale=0.4]{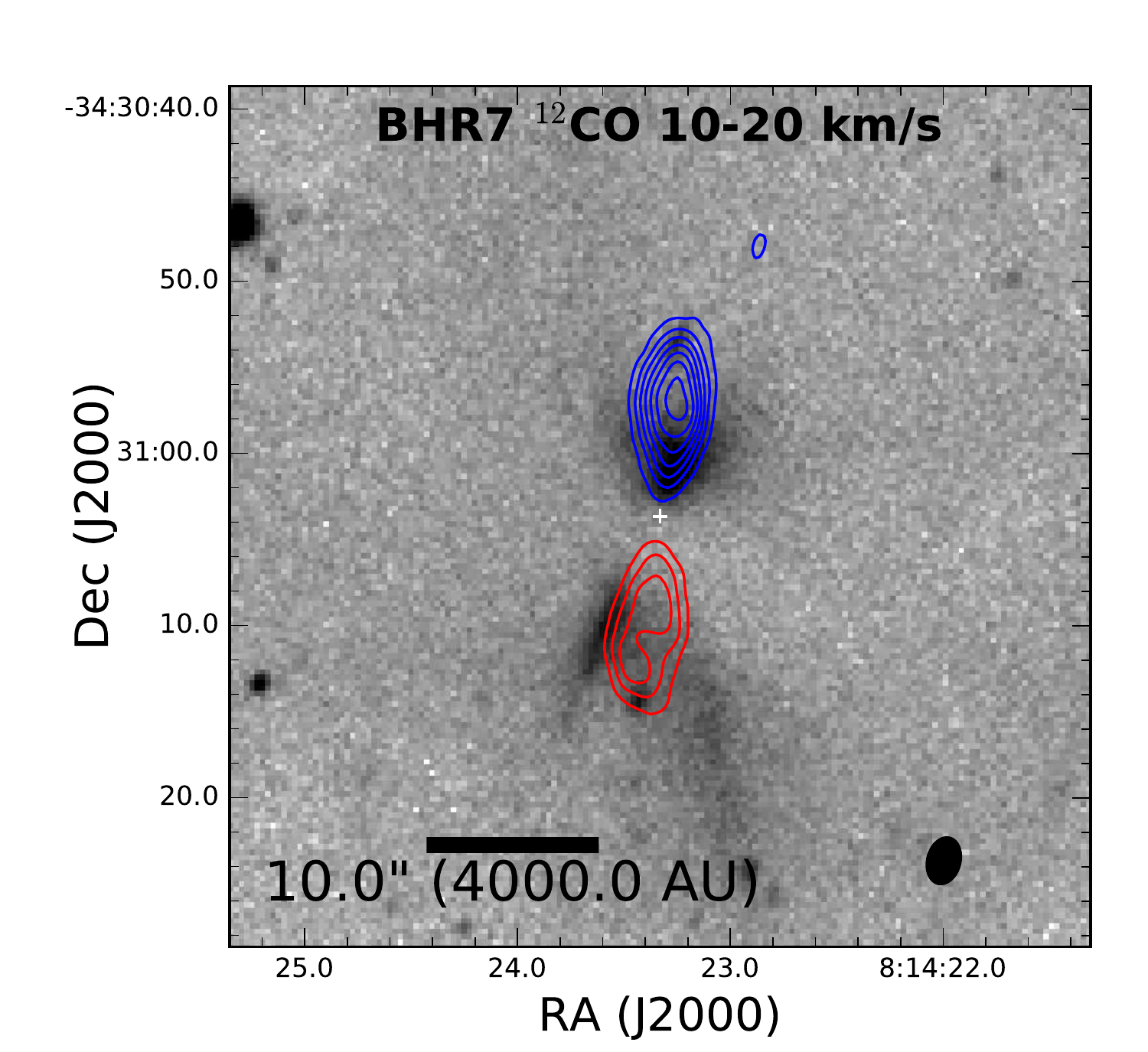}
\includegraphics[scale=0.4]{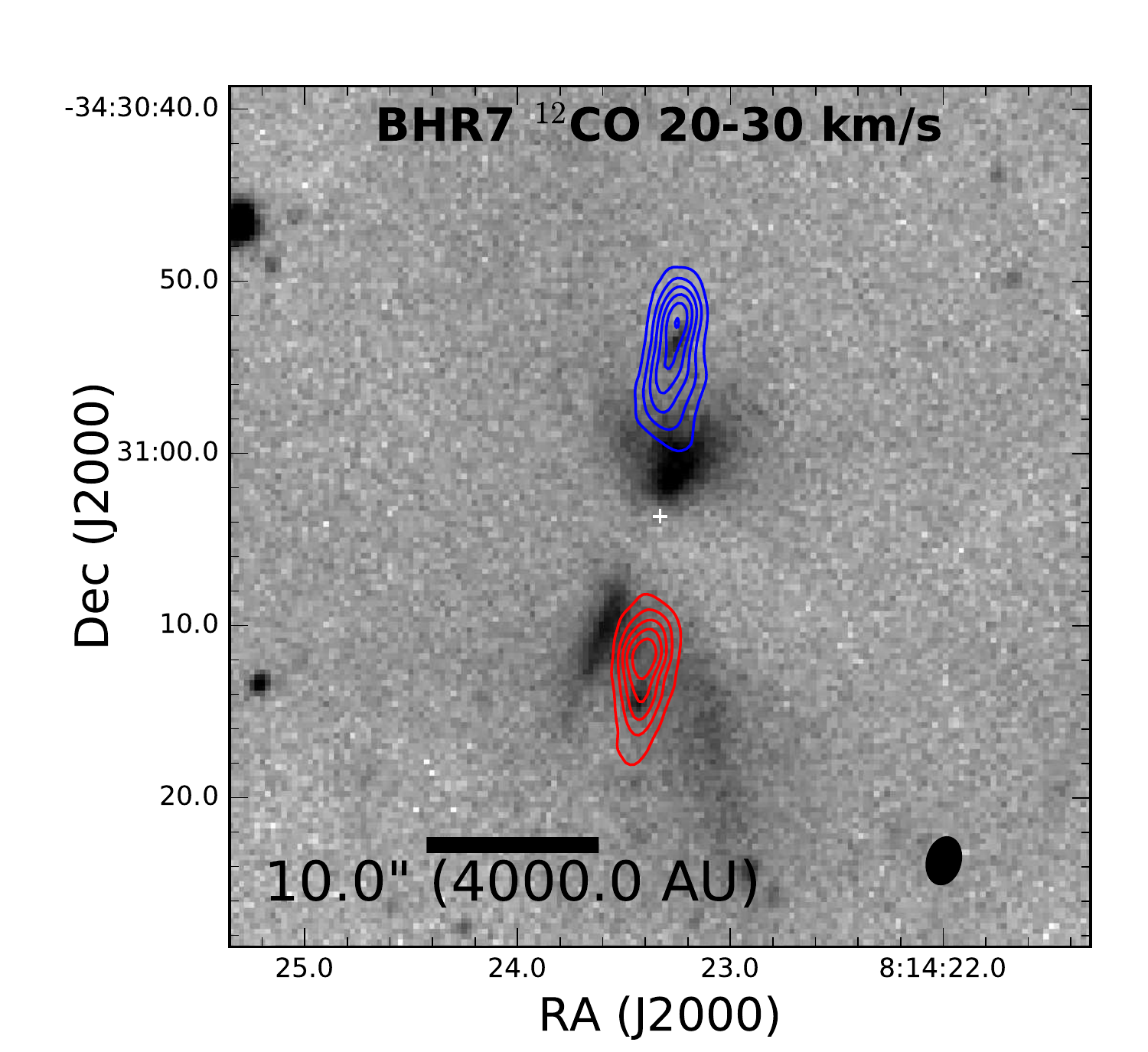}
\includegraphics[scale=0.4]{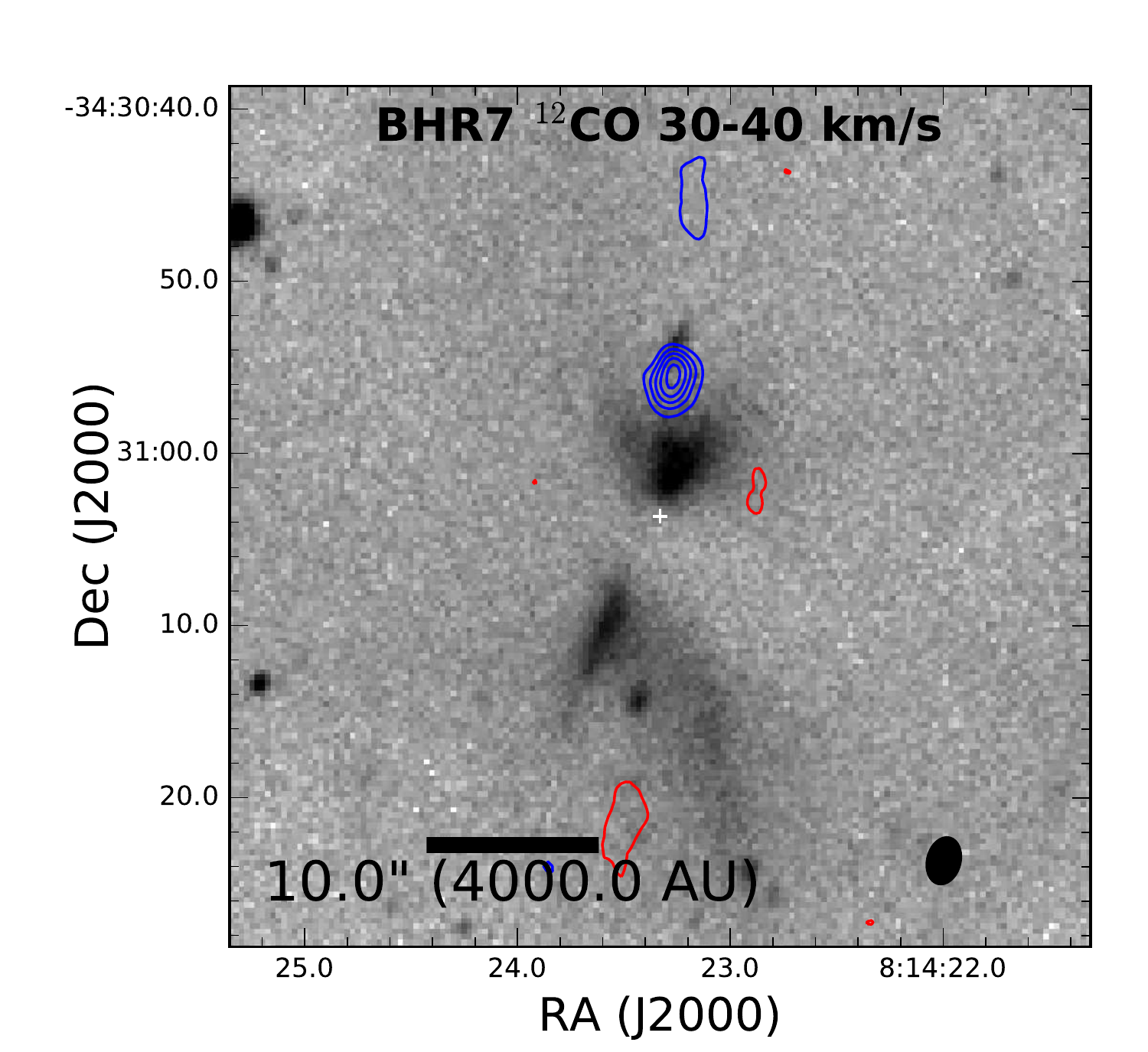}
\includegraphics[scale=0.35]{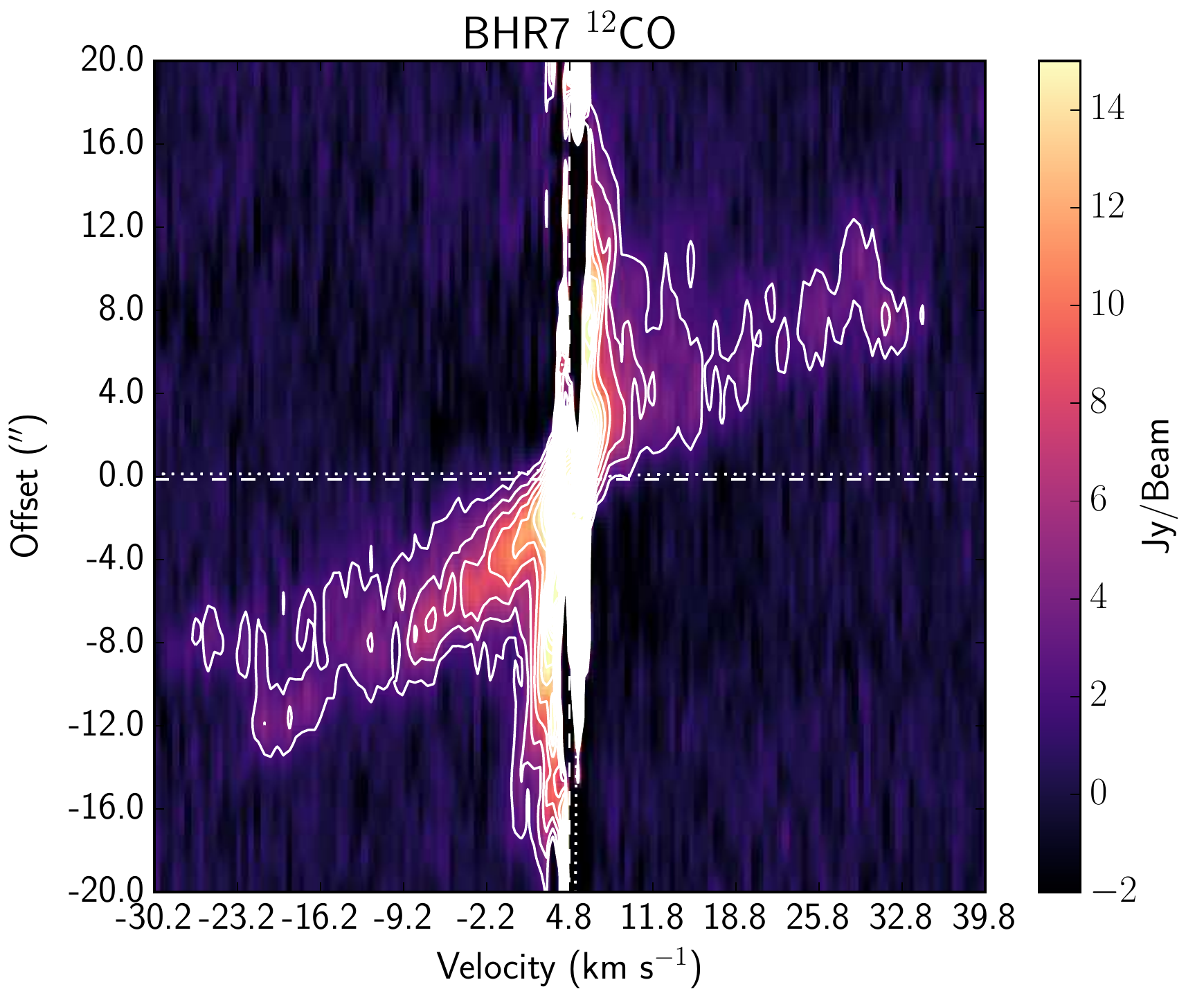}
\end{center}
\caption{SMA $^{12}$CO integrated intensity contours in multiple velocity ranges 
overlaid on the Ks-band near-infrared image. The outflow clearly shows evidence for
both a wide-angle (0$-$10~\kms) and a collimated high-velocity outflow ($>$10 \kms). The high-velocity
features also coincide with the knots of H$_2$ emission within the outflow cavity that are 
consistent with shocks. The bottom-most panel shows a position-velocity diagram extracted along the outflow axis, averages in an strip with an angular width of 10\farcs25.
 The beam is drawn in the lower right corner of each panel and is 2\farcs75$\times$1\farcs91. The contours in the 0$-$10~\kms\ panel start at 5.0$\sigma$ and increase in 7.5$\sigma$ intervals; $\sigma_{red}$=0.39~K and $\sigma_{blue}$=0.38~K~\kms. For the 10$-$20~\kms\ and 20$-$30~\kms\ panels, the contours
start at and increase at 5$\sigma$ where $\sigma_{red,blue}$=0.42~K~\kms. Lastly, for the 30$-$40~\kms\ panel
the contours start at 4$\sigma$ and increase at 2$\sigma$ intervals where $\sigma_{red,blue}$=0.42~K~\kms.}
\label{12co}
\end{figure}

While the near-infrared and \textit{Spitzer} data indicate the presence of an outflow carving out
cavities observed in scattered light, the SMA observations provide the first direct
measurements of the molecular outflow. The outflow from BHR7-MMS is clearly detected in $^{12}$CO 
and the integrated intensity maps of the red and blue-shifted sides of the outflow at different velocity ranges 
are shown in Figure \ref{12co}. The outflow exhibits structure typical for some Class 0 protostars, a wide-angle
component at low-velocity and a collimated high-velocity molecular jet \citep[e.g.,][]{hirano2010}. From the orientation of
the high-velocity jet, the outflow position angle (PA) is about -5.9\degr\ east of north. The PV diagram
taken along the outflow is also shown in Figure \ref{12co}. It shows the significant outflow emission out to $\pm$30~\kms\ from
the systemic velocity. 

The near-infrared and \textit{Spitzer} IRAC images also show two bright knots of 
shocked molecular line emission. Their peak
brightness at 4.5~\micron\ is indicative of the line emission being 
dominated by shock-excited H$_2$ emission. There is some
emission in $Ks$-band which also contains an H$_2$ line, but it 
is not as bright and/or peaked as the 4.5~\micron\ emission.
This could result from the $Ks$-band feature being more affected 
by extinction than the 4.5~\micron\ feature. The outflow
knots are located at nearly symmetric positions along the outflow axis. T
The northern knot is 11\farcs2 (4480~AU) from the continuum
source and the southern knot is 10\farcs75 (4300~AU) from the continuum source. 

The maximum observed velocity with respect to the systemic velocity is $\sim$30~\kms; this is 
approximate because there may be emission below our sensitivity limit at higher velocities. Correcting
this velocity for the approximate inclination of 73.2\degr\, the true velocity of the highest velocity
outflow is $\sim$104~\kms. If the H$_2$ knots are traveling along this outflow with the same velocity,
their time since ejection is $\sim$205~yr and would be moving at a 
rate of $\sim$22~AU~yr$^{-1}$ (0\farcs055~yr$^{-1}$).
Thus, the proper motion of these knots should be detectable now or in the near future given that the
images presented here were taken in 2008 (IRAC) and 2009 (JHKs).

We examined the outflow for indications of other shock-tracing species. 
We did detect SO emission, but 
it is centered toward the protostar position with a slight extension in the outflow direction
and only detected at low-velocities.
SiO was not detected in our observations. This indicates that the SO emission is not being produced 
by shocks in the 
extended
outflow and that the outflow shocks are not strong enough to produce detectable SiO emission.

\section{Discussion}

The data and results presented for BHR7 show a well-developed, isolated protostar system
that appears proto-typical in many ways. It has a well-defined outflow-carved cavity
that is apparent in scattered light between 1.6~\micron\ and 8~\micron, and the outflow
is also traced by CO emission, having a low-velocity wide-angle cavity and a narrow,
collimated higher-velocity flow that is aligned with the HH-knots apparent in the near-infrared.
The features of the outflow strongly resemble other `proto-type' systems like 
HH212 and L1448C \citep[e.g.,][]{zinnecker1998,lee2007,hirano2010}. However, it remains unclear
if shock-tracers like SiO are also associated with the jet; SiO was not detected in 
the current SMA data.

The velocity gradients, likely tracing rotation except at perhaps the largest scales, are exhibited in multiple molecular tracers (\ntdp, \cateo, \thco, and H$_2$CO). The \cateo\ and \thco\ appear to trace
to trace the rotating, infalling envelope while conserving angular momentum, as evidenced
by the R$^{-1}$ velocity profile shown in Figure \ref{peakpv}. Thus, it bears remarkable
similarity to the earlier data on L1527~IRS
from the SMA and Nobeyama Millimeter Array
\citep{ohashi1997, yen2013}, but 
BHR7-MMS 
has a stronger indication
of rotation. L1527~IRS, and other sources that also showed conserved angular momentum (and/or rotation)
in early data \citep[e.g., VLA~1623, HH212, Lupus MMS3][]{murillo2013b,lee2014, yen2015,yen2017}
were later found to have rotationally supported disks when viewed at higher resolution and sensitivity.

\subsection{Does BHR7-MMS Contain A Rotationally-Supported Disk?}

The \thco, \cateo, and H$_2$CO all show indications of rotation in the envelope
and the high-resolution \thco\ and \cateo, show rotation down to 100s of AU scales.
The velocity profile also shows conserved angular momentum from 1000s of AU down to 100s of AU.
The scales on which we observe rotation are 
small enough that magnetic braking will not be strong enough to reverse the increasing rotation
velocities \citep[e.g.,][]{zhao2017}. Nevertheless, the current spectral line data do not yet have resolution and
sensitivity to resolve and firmly identify the Keplerian region.

In addition to the strong rotation of the inner envelope, the SMA continuum observations at $\sim$0\farcs5
resolution show that the dust emission is resolved orthogonal to the outflow direction with a radius of
$\sim$104~AU (deconvolved Gaussian half-width at half maximum). 
The state at which we observe BHR7-MMS is remarkably similar to L1527~IRS and VLA 1623,
where both had resolved dust continuum orthogonal to the outflow and indications
of rotation from lower-resolution kinematic data. The analysis presented in Sections 3.5 and 3.7,
show that continuum data are consistent with 
the presence of a disk at the heart of the rotating envelope. 
However, we cannot conclusively conclude
that such a continuum structure is tracing a rotationally-supported disk around the protostar.

The molecular line kinematics do show rotation up to near the radius of the continuum emission.
Thus, if the disk is rotationally-supported, we can simplistically estimate 
the protostar mass by assuming a Keplerian radius. We adopt a Keplerian radius
of 120~AU from resolved dust continuum
emission discussed in Section 3.5.  This radius corresponds to the radius of the
3$\sigma$ contour and is slightly larger than the deconvolved half-width at half maximum from the
Gaussian fit. For a Keplerian radius of 120~AU, we extrapolated the
conserved angular momentum power-law, finding a velocity of 2.58~\kms\ at the adopted
Keplerian radius. We show in Figure \ref{peakpv} that this radius is just beyond the smallest radius detected
by the \thco\ and \cateo\ data. 
Using the formula v~=~(GM/R)$^{0.5}$, we find that with 
the assumptions outlined above, the protostar mass could be 1.0~$M_{\sun}$.
This estimate incorporates the 
$\sim$73\degr\ inclination calculated in Section 3.5 from the Gaussian fit to the continuum

This estimate of central protostar mass is contingent upon both the continuum tracing a fully Keplerian disk 
(which may not be the case). If the true Keplerian disk is smaller, then the protostar mass would be larger;
the protostar mass will scale inversely with the Keplerian radius. For example, if the Keplerian radius is actually
60 AU, then the protostar mass would be $\sim$2~M$_{\sun}$. While there is considerable uncertainty in the
protostar mass, estimates substantially above $\sim$1~M$_{\sun}$ may be overly large 
considering the bolometric luminosity of 9.3~L$_{\sun}$.

\subsection{Likelihood of Gravitational Instability}

The mass of the 
postulated
disk was found to be large from both the isothermal dust mass 
calculation ($M_{disk}$=0.42~$M_{\sun}$) and the visibility amplitude profile
analysis ($M_{disk}$=0.47~$M_{\sun}$) in Sections 3.5 and 3.7. However, both of these
mass estimates assume a gas to dust mass ratio of 100:1, which can evolve \citep{ansdell2016}, 
but it is unclear if any reduction of this ratio is expected in the protostellar phase. Assuming
that both the $M_{disk}$=0.42~$M_{\sun}$ value and $M_{*}$=1.0~$M_{\sun}$ are accurate, we can examine
the likelihood of gravitational instability in the disk of BHR7-MMS. 

The typical analytic criterion for stability of a rotating system is given by Toomre's
Q, calculating the ratio of thermal pressure support and rotation shear to gravity in the form 
\begin{equation}
Q = \frac{c_s\Omega}{\pi\Sigma G}.
\end{equation}
Values greater than 1 indicate that the disk is stable against self-gravity and values less than 1 indicate that the disk 
is unstable and may be prone to fragmentation.

This equation can be rewritten for a disk \citep{kratter2016,tobin2016} in the form
\begin{equation}
\label{eq:qapprox}
Q \approx 2\frac{M_*}{M_d}\frac{H}{r}.
\end{equation}
$H=c_s/\Omega$ is the disk scale height, $c_s$ the disk sound speed, and $\Omega$ the Keplerian angular velocity. 
Using the disk and star masses derived above, and a 
typical disk temperature of $30$~K ($c_s$~=~0.3~\kms) at a 
radius of 50~AU ($\Omega$~$\sim$~5.6$\times$10$^{-10}$~s$^{-1}$), $Q\approx 0.4$. We can also see
from this equation that Q will scale as M$_{*}^{0.5}$ 
from the angular velocity dependence on protostar mass,
therefore if the current disk mass
is accurate, it is possible that the disk is gravitationally unstable. However, another limiting
factor may be the uncertainty in the distance ($D$) to this protostar because the disk mass
from dust emission will scale as $D^{2}$, but protostar mass measured from the spectral line 
emission will scale linearly with distance, making Q scale as $D^{1.5}$. Thus, if BHR7 is somehow a factor of 2 closer than
the adopted 400~pc distance, the disk mass will be 4$\times$ lower
and the protostar mass will be 0.5~$M_{\sun}$, making Q $\sim$ 1.1. We also note that it
is possible that we are underestimating the mass in the disk component if it is optically thick.

Future high-resolution
observations will be required to more accurately determine the disk and protostar mass
to better assess its likelihood of gravitational instability. Moreover, higher-resolution
continuum imaging may directly reveal fragments or spiral structure that
has recently been uncovered in disks by ALMA \citep{tobin2016,perez2016}. Finally, Gaia
parallaxes of foreground/background stars will enable the distance of the BHR7 globule to
be better constrained.

\section{Conclusions}

We have characterized the protostar within the isolated dark globule BHR7 from the near-infrared
to the submillimeter using data from the CTIO 4m, \textit{Spitzer}, \textit{IRAS},
 \textit{Herschel}, SEST SIMBA, and the SMA. We have made the first determination of bolometric
luminosity 9.3~L$_{\sun}$ and bolometric temperature (50.5~K) for this protostar, confirming that
it is a Class 0 protostar with a luminosity that is higher than the $\sim$1~L$_{\sun}$ median in the nearby star-forming regions \citep{dunham2014}. The near-infrared at Ks-band
and \textit{Spitzer} IRAC bands show that the source 
has a well-defined
outflow cavity separated by a dark lane. The outflow
cavity is traced by an impressive CO outflow that shows both low-velocity
wide-angle emission and a collimated, high-velocity jet that has an axis
consistent with observed knots of shocked-excited H$_2$ gas.

We conducted further submillimeter interferometric observations using the 
SMA in three configurations to examine both the molecular line emission from the 
envelope down to near the scale of the disk, as well as the small-scale continuum structure.
The \thco\ and \cateo\ kinematics indicate clear rotation in the inner envelope and
the rotation curve in the the inner envelope is rotating consistent
with conserved angular momentum.
The SMA continuum observations with $\sim$0\farcs5 resolution resolve the continuum
orthogonal to the outflow direction, possibly tracing a disk. The deconvolved
size of the continuum indicates that the disk radius may be $\sim$100~AU
and have an inclination near 73\degr. Analysis of the visibility amplitude profile
down to sub-arcsecond scales requires the presence of a compact structure other than
a power-law or rotationally-flattened envelope. A disk plus a rotationally-flattened
envelope can reproduce the visibility amplitude profile
The strong envelope rotation, coupled with the
resolved continuum image and visibility amplitude profile
are possible evidence for a rotationally-supported
disk within this system. We
estimate the disk mass from the continuum emission to be $\sim$0.32~M$_{\sun}$ and the 
gas kinematics (assuming a rotationally supported disk radius of $\sim$120~AU) indicate
that the protostar mass could be $\sim$1~M$_{\sun}$. With these two pieces of information,
we calculate Toomre's Q, finding that
if the observations reflect a rotationally-supported
disk, it
may be gravitationally unstable.

Furthermore, H$_2$CO and \ntdp\ observations 
from larger radii indicate that the envelope still exhibits a velocity gradient beyond
1000~AU with a gradient direction consistent with the smaller-scale \thco\ and \cateo\
emission; this possibly indicates bulk rotation of the outer envelope.

These observations as a whole demonstrate that BHR7 may be an excellent proto-type
source for studies of star formation, given its isolation from confusing, external
influence and its classic outflow and inner envelope rotation profile. Future observations
will reveal whether or not there is 
indeed a rotationally-supported disk and if it
is currently only forming a single star. However, despite its gifts, the ambiguity
of the distance toward BHR7 must be resolved to more firmly characterize its properties, and
forthcoming parallax data from Gaia should enable more firm constraints on its distance in the near
future.

J.J.T. is acknowledges support from the Homer L. Dodge Endowed Chair
and grant 639.041.439 from the Netherlands
Organisation for Scientific Research (NWO).
This work is based in part on observations 
made with Herschel, a European Space Agency Cornerstone Mission with significant 
participation by NASA. The authors wish to thank the SMA staff
for their tireless efforts in keeping the facility operational and 
conducting the observations. The Submillimeter Array is a joint 
project between the Smithsonian Astrophysical Observatory and the Academia Sinica 
Institute of Astronomy and Astrophysics and is funded by the Smithsonian 
Institution and the Academia Sinica. The authors wish to recognize and acknowledge the very significant cultural role and reverence that the summit of Mauna Kea has always had within the indigenous Hawaiian community.  We are most fortunate to have the opportunity to conduct observations from this mountain.

{\it Facilities:}  \facility{SMA}, \facility{\textit{Herschel}}, \facility{\textit{Spitzer}}, \facility{Blanco (ISPI)}

\begin{small}
\bibliographystyle{apj}
\bibliography{ms}
\end{small}

\clearpage

\clearpage

\clearpage

\input{tab1}
\input{tab2}
\end{document}

%% file: tab1.tex
\begin{deluxetable}{llllll}
\tablewidth{0pt}
\rotate
\tabletypesize{\scriptsize}
\tablecaption{SMA Observations}
\tablehead{
  \colhead{Source} & \colhead{RA} & \colhead{Dec}  &\colhead{Config.\tablenotemark{a}}  & \colhead{Date} & \colhead{Calibrators}\\
             & \colhead{(J2000)} &  \colhead{(J2000)}    &  &    \colhead{(UT)}          &   \colhead{(Gain, Flux)}   \\
}
\startdata
BHR7             & 8:14:23.326  & -34:31:05.7  & VEX & 28 Jan 2015 & 0747-311, Callisto    \\
BHR7             & 8:14:23.33  & -34:31:03.7  & COMP & 25 Dec 2015 &  0747-331, Uranus  \\
BHR7             & 8:14:23.33  & -34:31:03.7  & EXT & 02 Apr 2016 & 0747-311, Ganymede    \\
\enddata
\tablecomments{The position listed for VEX corresponds to the phase center of the observations, not
the coordinates of the continuum emission, and the positions for EXT and COMP are centered on the continuum
source.}
\tablenotemark{a}{VEX - Very Extended, EXT - Extended, and COMP - Compact} 
\end{deluxetable}

%% file: tab2.tex
\begin{deluxetable}{lll}
\tablewidth{0pc}
\tabletypesize{\scriptsize}
\tablecaption{BHR7 Photometry}
\tablehead{
  \colhead{Wavelength}   & \colhead{Flux Density}  & \colhead{Reference}\\
  \colhead{(\micron)} & \colhead{(Jy)}   
}
\startdata
1.25 &  0.00087 $\pm$ 0.00008 & 1\\
1.6  &  0.0022 $\pm$ 0.0002 & 1\\
2.15  &  0.0054 $\pm $0.0005 & 1\\
3.6 &   0.031 $\pm$ 0.001 & 1\\
4.5 &   0.066 $\pm$ 0.002 & 1\\
5.8 &   0.044 $\pm$ 0.002 & 1\\
8.0 &   0.027 $\pm$ 0.003 & 1\\
25.0 & 0.296 $\pm$ 0.03 & IRAS\\
60.0 & 11.9 $\pm$ 1.9 & IRAS\\
100.0 & 40.5 $\pm$ 4.5 & IRAS\\

250.0 & 34.56 $\pm$ 1.4 & 1\\
350.0 & 19.95 $\pm$ 0.84 & 1\\
500.0 & 9.78 $\pm$ 0.47 & 1\\

1200.0 & 0.76 $\pm$ 0.15  & 1\\

\enddata
\tablecomments{
\textbf{
The uncertainties listed are statistical only and do not include systematic uncertainty from the absolute
flux calibration accuracy. Furthermore, emission in the SPIRE bands is extended beyond the radius for which
we measure it.
}
 References: (1) This work.} 
\end{deluxetable}

%% file: ms.bbl
\begin{thebibliography}{}
\expandafter\ifx\csname natexlab\endcsname\relax\def\natexlab#1{#1}\fi

\bibitem[{{Andr\'e} {et~al.}(1993){Andr\'e}, {Ward-Thompson}, \&
  {Barsony}}]{andre1993}
{Andr\'e}, P., {Ward-Thompson}, D., \& {Barsony}, M. 1993, \apj, 406, 122

\bibitem[{{Ansdell} {et~al.}(2016){Ansdell}, {Williams}, {van der Marel},
  {Carpenter}, {Guidi}, {Hogerheijde}, {Mathews}, {Manara}, {Miotello},
  {Natta}, {Oliveira}, {Tazzari}, {Testi}, {van Dishoeck}, \& {van
  Terwisga}}]{ansdell2016}
{Ansdell}, M., {Williams}, J.~P., {van der Marel}, N., {et~al.} 2016, \apj,
  828, 46

\bibitem[{{Aso} {et~al.}(2017){Aso}, {Ohashi}, {Aikawa}, {Machida}, {Saigo},
  {Saito}, {Takakuwa}, {Tomida}, {Tomisaka}, \& {Yen}}]{aso2017}
{Aso}, Y., {Ohashi}, N., {Aikawa}, Y., {et~al.} 2017, ArXiv e-prints,
  arXiv:1707.08697

\bibitem[{{Bergin} {et~al.}(2001){Bergin}, {Ciardi}, {Lada}, {Alves}, \&
  {Lada}}]{bergin2001}
{Bergin}, E.~A., {Ciardi}, D.~R., {Lada}, C.~J., {Alves}, J., \& {Lada}, E.~A.
  2001, \apj, 557, 209

\bibitem[{{Bohlin} {et~al.}(1978){Bohlin}, {Savage}, \& {Drake}}]{bohlin1978}
{Bohlin}, R.~C., {Savage}, B.~D., \& {Drake}, J.~F. 1978, \apj, 224, 132

\bibitem[{{Bourke} {et~al.}(1995{\natexlab{a}}){Bourke}, {Hyland}, \&
  {Robinson}}]{bourke1995a}
{Bourke}, T.~L., {Hyland}, A.~R., \& {Robinson}, G. 1995{\natexlab{a}}, \mnras,
  276, 1052

\bibitem[{{Bourke} {et~al.}(1995{\natexlab{b}}){Bourke}, {Hyland}, {Robinson},
  {James}, \& {Wright}}]{bourke1995b}
{Bourke}, T.~L., {Hyland}, A.~R., {Robinson}, G., {James}, S.~D., \& {Wright},
  C.~M. 1995{\natexlab{b}}, \mnras, 276, 1067

\bibitem[{{Caselli} {et~al.}(1999){Caselli}, {Walmsley}, {Tafalla}, {Dore}, \&
  {Myers}}]{caselli1999}
{Caselli}, P., {Walmsley}, C.~M., {Tafalla}, M., {Dore}, L., \& {Myers}, P.~C.
  1999, \apjl, 523, L165

\bibitem[{{Cassen} \& {Moosman}(1981)}]{cassen1981}
{Cassen}, P., \& {Moosman}, A. 1981, \icarus, 48, 353

\bibitem[{{Chen} {et~al.}(1995){Chen}, {Myers}, {Ladd}, \& {Wood}}]{chen1995}
{Chen}, H., {Myers}, P.~C., {Ladd}, E.~F., \& {Wood}, D.~O.~S. 1995, \apj, 445,
  377

\bibitem[{{Cieza} {et~al.}(2016){Cieza}, {Casassus}, {Tobin}, {Bos},
  {Williams}, {Perez}, {Zhu}, {Caceres}, {Canovas}, {Dunham}, {Hales},
  {Prieto}, {Principe}, {Schreiber}, {Ruiz-Rodriguez}, \& {Zurlo}}]{cieza2016}
{Cieza}, L.~A., {Casassus}, S., {Tobin}, J., {et~al.} 2016, \nat, 535, 258

\bibitem[{{Codella} {et~al.}(2014){Codella}, {Cabrit}, {Gueth}, {Podio},
  {Leurini}, {Bachiller}, {Gusdorf}, {Lefloch}, {Nisini}, {Tafalla}, \&
  {Yvart}}]{codella2014b}
{Codella}, C., {Cabrit}, S., {Gueth}, F., {et~al.} 2014, \aap, 568, L5

\bibitem[{{Crapsi} {et~al.}(2004){Crapsi}, {Caselli}, {Walmsley}, {Tafalla},
  {Lee}, {Bourke}, \& {Myers}}]{crapsi2004}
{Crapsi}, A., {Caselli}, P., {Walmsley}, C.~M., {et~al.} 2004, \aap, 420, 957

\bibitem[{{Dunham} {et~al.}(2014){Dunham}, {Stutz}, {Allen}, {Evans},
  {Fischer}, {Megeath}, {Myers}, {Offner}, {Poteet}, {Tobin}, \&
  {Vorobyov}}]{dunham2014}
{Dunham}, M.~M., {Stutz}, A.~M., {Allen}, L.~E., {et~al.} 2014, Protostars and
  Planets VI, 195

\bibitem[{{Emprechtinger} {et~al.}(2009){Emprechtinger}, {Caselli}, {Volgenau},
  {Stutzki}, \& {Wiedner}}]{emprechtinger2009}
{Emprechtinger}, M., {Caselli}, P., {Volgenau}, N.~H., {Stutzki}, J., \&
  {Wiedner}, M.~C. 2009, \aap, 493, 89

\bibitem[{{Evans} {et~al.}(2003){Evans}, {Allen}, {Blake}, {Boogert}, {Bourke},
  {Harvey}, {Kessler}, {Koerner}, {Lee}, {Mundy}, {Myers}, {Padgett},
  {Pontoppidan}, {Sargent}, {Stapelfeldt}, {van Dishoeck}, {Young}, \&
  {Young}}]{evans2003}
{Evans}, II, N.~J., {Allen}, L.~E., {Blake}, G.~A., {et~al.} 2003, \pasp, 115,
  965

\bibitem[{{Fazio} {et~al.}(2004){Fazio}, {Hora}, {Allen}, {Ashby}, {Barmby},
  {Deutsch}, {Huang}, {Kleiner}, {Marengo}, {Megeath}, {Melnick}, {Pahre},
  {Patten}, {Polizotti}, {Smith}, {Taylor}, {Wang}, {Willner}, {Hoffmann},
  {Pipher}, {Forrest}, {McMurty}, {McCreight}, {McKelvey}, {McMurray}, {Koch},
  {Moseley}, {Arendt}, {Mentzell}, {Marx}, {Losch}, {Mayman}, {Eichhorn},
  {Krebs}, {Jhabvala}, {Gezari}, {Fixsen}, {Flores}, {Shakoorzadeh}, {Jungo},
  {Hakun}, {Workman}, {Karpati}, {Kichak}, {Whitley}, {Mann}, {Tollestrup},
  {Eisenhardt}, {Stern}, {Gorjian}, {Bhattacharya}, {Carey}, {Nelson},
  {Glaccum}, {Lacy}, {Lowrance}, {Laine}, {Reach}, {Stauffer}, {Surace},
  {Wilson}, {Wright}, {Hoffman}, {Domingo}, \& {Cohen}}]{fazio2004}
{Fazio}, G.~G., {Hora}, J.~L., {Allen}, L.~E., {et~al.} 2004, \apjs, 154, 10

\bibitem[{{Fischer} {et~al.}(2017){Fischer}, {Megeath}, {Furlan}, {Ali},
  {Stutz}, {Tobin}, {Osorio}, {Stanke}, {Manoj}, {Poteet}, {Booker},
  {Hartmann}, {Wilson}, {Myers}, \& {Watson}}]{fischer2017}
{Fischer}, W.~J., {Megeath}, S.~T., {Furlan}, E., {et~al.} 2017, \apj, 840, 69

\bibitem[{{Furlan} {et~al.}(2016){Furlan}, {Fischer}, {Ali}, {Stutz}, {Stanke},
  {Tobin}, {Megeath}, {Osorio}, {Hartmann}, {Calvet}, {Poteet}, {Booker},
  {Manoj}, {Watson}, \& {Allen}}]{furlan2016}
{Furlan}, E., {Fischer}, W.~J., {Ali}, B., {et~al.} 2016, \apjs, 224, 5

\bibitem[{{Griffin} {et~al.}(2010){Griffin}, {Abergel}, {Abreu}, {Ade},
  {Andr{\'e}}, {Augueres}, {Babbedge}, {Bae}, {Baillie}, {Baluteau}, {Barlow},
  {Bendo}, {Benielli}, {Bock}, {Bonhomme}, {Brisbin}, {Brockley-Blatt},
  {Caldwell}, {Cara}, {Castro-Rodriguez}, {Cerulli}, {Chanial}, {Chen},
  {Clark}, {Clements}, {Clerc}, {Coker}, {Communal}, {Conversi}, {Cox},
  {Crumb}, {Cunningham}, {Daly}, {Davis}, {de Antoni}, {Delderfield}, {Devin},
  {di Giorgio}, {Didschuns}, {Dohlen}, {Donati}, {Dowell}, {Dowell}, {Duband},
  {Dumaye}, {Emery}, {Ferlet}, {Ferrand}, {Fontignie}, {Fox}, {Franceschini},
  {Frerking}, {Fulton}, {Garcia}, {Gastaud}, {Gear}, {Glenn}, {Goizel},
  {Griffin}, {Grundy}, {Guest}, {Guillemet}, {Hargrave}, {Harwit}, {Hastings},
  {Hatziminaoglou}, {Herman}, {Hinde}, {Hristov}, {Huang}, {Imhof}, {Isaak},
  {Israelsson}, {Ivison}, {Jennings}, {Kiernan}, {King}, {Lange}, {Latter},
  {Laurent}, {Laurent}, {Leeks}, {Lellouch}, {Levenson}, {Li}, {Li},
  {Lilienthal}, {Lim}, {Liu}, {Lu}, {Madden}, {Mainetti}, {Marliani}, {McKay},
  {Mercier}, {Molinari}, {Morris}, {Moseley}, {Mulder}, {Mur}, {Naylor},
  {Nguyen}, {O'Halloran}, {Oliver}, {Olofsson}, {Olofsson}, {Orfei}, {Page},
  {Pain}, {Panuzzo}, {Papageorgiou}, {Parks}, {Parr-Burman}, {Pearce},
  {Pearson}, {P{\'e}rez-Fournon}, {Pinsard}, {Pisano}, {Podosek}, {Pohlen},
  {Polehampton}, {Pouliquen}, {Rigopoulou}, {Rizzo}, {Roseboom}, {Roussel},
  {Rowan-Robinson}, {Rownd}, {Saraceno}, {Sauvage}, {Savage}, {Savini},
  {Sawyer}, {Scharmberg}, {Schmitt}, {Schneider}, {Schulz}, {Schwartz},
  {Shafer}, {Shupe}, {Sibthorpe}, {Sidher}, {Smith}, {Smith}, {Smith},
  {Spencer}, {Stobie}, {Sudiwala}, {Sukhatme}, {Surace}, {Stevens}, {Swinyard},
  {Trichas}, {Tourette}, {Triou}, {Tseng}, {Tucker}, {Turner}, {Vaccari},
  {Valtchanov}, {Vigroux}, {Virique}, {Voellmer}, {Walker}, {Ward}, {Waskett},
  {Weilert}, {Wesson}, {White}, {Whitehouse}, {Wilson}, {Winter}, {Woodcraft},
  {Wright}, {Xu}, {Zavagno}, {Zemcov}, {Zhang}, \& {Zonca}}]{griffin2010}
{Griffin}, M.~J., {Abergel}, A., {Abreu}, A., {et~al.} 2010, \aap, 518, L3

\bibitem[{{Hartley} {et~al.}(1986){Hartley}, {Tritton}, {Manchester}, {Smith},
  \& {Goss}}]{hartley1986}
{Hartley}, M., {Tritton}, S.~B., {Manchester}, R.~N., {Smith}, R.~M., \&
  {Goss}, W.~M. 1986, \aaps, 63, 27

\bibitem[{{Hirano} {et~al.}(2010){Hirano}, {Ho}, {Liu}, {Shang}, {Lee}, \&
  {Bourke}}]{hirano2010}
{Hirano}, N., {Ho}, P.~P.~T., {Liu}, S.-Y., {et~al.} 2010, \apj, 717, 58

\bibitem[{{Ho} {et~al.}(2004){Ho}, {Moran}, \& {Lo}}]{ho2004}
{Ho}, P.~T.~P., {Moran}, J.~M., \& {Lo}, K.~Y. 2004, \apjl, 616, L1

\bibitem[{{J{\o}rgensen}(2004)}]{jorgensen2004}
{J{\o}rgensen}, J.~K. 2004, \aap, 424, 589

\bibitem[{{K\"{a}mpgen}(2002)}]{kampgen2002}
{K\"{a}mpgen}, K. 2002, Master's~Thesis

\bibitem[{{K{\"a}mpgen} {et~al.}(2004){K{\"a}mpgen}, {Chini}, {Nielbock}, \&
  {Albrecht}}]{kampgen2004}
{K{\"a}mpgen}, K., {Chini}, R., {Nielbock}, M., \& {Albrecht}, M. 2004, in The
  Dense Interstellar Medium in Galaxies, ed. S.~{Pfalzner}, C.~{Kramer},
  C.~{Staubmeier}, \& A.~{Heithausen}, Vol.~91, 405

\bibitem[{{Kratter} \& {Lodato}(2016)}]{kratter2016}
{Kratter}, K., \& {Lodato}, G. 2016, \araa, 54, 271

\bibitem[{{Kratter} {et~al.}(2010){Kratter}, {Matzner}, {Krumholz}, \&
  {Klein}}]{kratter2010}
{Kratter}, K.~M., {Matzner}, C.~D., {Krumholz}, M.~R., \& {Klein}, R.~I. 2010,
  \apj, 708, 1585

\bibitem[{{Langer}(1985)}]{langer1985}
{Langer}, W.~D. 1985, in Protostars and Planets II, ed. D.~C. {Black} \& M.~S.
  {Matthews}, 650--667

\bibitem[{{Lee} {et~al.}(2014){Lee}, {Hirano}, {Zhang}, {Shang}, {Ho}, \&
  {Krasnopolsky}}]{lee2014}
{Lee}, C.-F., {Hirano}, N., {Zhang}, Q., {et~al.} 2014, \apj, 786, 114

\bibitem[{{Lee} {et~al.}(2007){Lee}, {Ho}, {Hirano}, {Beuther}, {Bourke},
  {Shang}, \& {Zhang}}]{lee2007}
{Lee}, C.-F., {Ho}, P.~T.~P., {Hirano}, N., {et~al.} 2007, \apj, 659, 499

\bibitem[{{Lindberg} {et~al.}(2014){Lindberg}, {J{\o}rgensen}, {Brinch},
  {Haugb{\o}lle}, {Bergin}, {Harsono}, {Persson}, {Visser}, \&
  {Yamamoto}}]{lindberg2014}
{Lindberg}, J.~E., {J{\o}rgensen}, J.~K., {Brinch}, C., {et~al.} 2014, \aap,
  566, A74

\bibitem[{{Looney} {et~al.}(2007){Looney}, {Tobin}, \& {Kwon}}]{looney2007}
{Looney}, L.~W., {Tobin}, J.~J., \& {Kwon}, W. 2007, \apjl, 670, L131

\bibitem[{{McMullin} {et~al.}(2007){McMullin}, {Waters}, {Schiebel}, {Young},
  \& {Golap}}]{mcmullin2007}
{McMullin}, J.~P., {Waters}, B., {Schiebel}, D., {Young}, W., \& {Golap}, K.
  2007, in Astronomical Society of the Pacific Conference Series, Vol. 376,
  Astronomical Data Analysis Software and Systems XVI, ed. R.~A. {Shaw},
  F.~{Hill}, \& D.~J. {Bell}, 127

\bibitem[{{Murillo} \& {Lai}(2013)}]{murillo2013}
{Murillo}, N.~M., \& {Lai}, S.-P. 2013, \apjl, 764, L15

\bibitem[{{Murillo} {et~al.}(2013){Murillo}, {Lai}, {Bruderer}, {Harsono}, \&
  {van Dishoeck}}]{murillo2013b}
{Murillo}, N.~M., {Lai}, S.-P., {Bruderer}, S., {Harsono}, D., \& {van
  Dishoeck}, E.~F. 2013, \aap, 560, A103

\bibitem[{{Ohashi} {et~al.}(1997){Ohashi}, {Hayashi}, {Ho}, \&
  {Momose}}]{ohashi1997}
{Ohashi}, N., {Hayashi}, M., {Ho}, P.~T.~P., \& {Momose}, M. 1997, \apj, 475,
  211

\bibitem[{{Ohashi} {et~al.}(2014){Ohashi}, {Saigo}, {Aso}, {Aikawa},
  {Koyamatsu}, {Machida}, {Saito}, {Takahashi}, {Takakuwa}, {Tomida},
  {Tomisaka}, \& {Yen}}]{ohashi2014}
{Ohashi}, N., {Saigo}, K., {Aso}, Y., {et~al.} 2014, \apj, 796, 131

\bibitem[{{Ossenkopf} \& {Henning}(1994)}]{ossenkopf1994}
{Ossenkopf}, V., \& {Henning}, T. 1994, \aap, 291, 943

\bibitem[{{P{\'e}rez} {et~al.}(2016){P{\'e}rez}, {Carpenter}, {Andrews},
  {Ricci}, {Isella}, {Linz}, {Sargent}, {Wilner}, {Henning}, {Deller},
  {Chandler}, {Dullemond}, {Lazio}, {Menten}, {Corder}, {Storm}, {Testi},
  {Tazzari}, {Kwon}, {Calvet}, {Greaves}, {Harris}, \& {Mundy}}]{perez2016}
{P{\'e}rez}, L.~M., {Carpenter}, J.~M., {Andrews}, S.~M., {et~al.} 2016,
  Science, 353, 1519

\bibitem[{{Poglitsch} {et~al.}(2010){Poglitsch}, {Waelkens}, {Geis},
  {Feuchtgruber}, {Vandenbussche}, {Rodriguez}, {Krause}, {Renotte}, {van
  Hoof}, {Saraceno}, {Cepa}, {Kerschbaum}, {Agn{\`e}se}, {Ali}, {Altieri},
  {Andreani}, {Augueres}, {Balog}, {Barl}, {Bauer}, {Belbachir}, {Benedettini},
  {Billot}, {Boulade}, {Bischof}, {Blommaert}, {Callut}, {Cara}, {Cerulli},
  {Cesarsky}, {Contursi}, {Creten}, {De Meester}, {Doublier}, {Doumayrou},
  {Duband}, {Exter}, {Genzel}, {Gillis}, {Gr{\"o}zinger}, {Henning},
  {Herreros}, {Huygen}, {Inguscio}, {Jakob}, {Jamar}, {Jean}, {de Jong},
  {Katterloher}, {Kiss}, {Klaas}, {Lemke}, {Lutz}, {Madden}, {Marquet},
  {Martignac}, {Mazy}, {Merken}, {Montfort}, {Morbidelli}, {M{\"u}ller},
  {Nielbock}, {Okumura}, {Orfei}, {Ottensamer}, {Pezzuto}, {Popesso},
  {Putzeys}, {Regibo}, {Reveret}, {Royer}, {Sauvage}, {Schreiber}, {Stegmaier},
  {Schmitt}, {Schubert}, {Sturm}, {Thiel}, {Tofani}, {Vavrek}, {Wetzstein},
  {Wieprecht}, \& {Wiezorrek}}]{poglitsch2010}
{Poglitsch}, A., {Waelkens}, C., {Geis}, N., {et~al.} 2010, \aap, 518, L2

\bibitem[{{Robitaille}(2011)}]{robitaille2011}
{Robitaille}, T.~P. 2011, \aap, 536, A79

\bibitem[{{Sakai} {et~al.}(2014){Sakai}, {Oya}, {Sakai}, {Watanabe}, {Hirota},
  {Ceccarelli}, {Kahane}, {Lopez-Sepulcre}, {Lefloch}, {Vastel}, {Bottinelli},
  {Caux}, {Coutens}, {Aikawa}, {Takakuwa}, {Ohashi}, {Yen}, \&
  {Yamamoto}}]{sakai2014}
{Sakai}, N., {Oya}, Y., {Sakai}, T., {et~al.} 2014, \apjl, 791, L38

\bibitem[{{Santos} {et~al.}(1998){Santos}, {Yun}, {Santos}, \&
  {Marreiros}}]{santos1998}
{Santos}, N.~C., {Yun}, J.~L., {Santos}, C.~A., \& {Marreiros}, R.~G. 1998,
  \aj, 116, 1376

\bibitem[{{Sault} {et~al.}(1995){Sault}, {Teuben}, \& {Wright}}]{sault1995}
{Sault}, R.~J., {Teuben}, P.~J., \& {Wright}, M.~C.~H. 1995, in Astronomical
  Society of the Pacific Conference Series, Vol.~77, Astronomical Data Analysis
  Software and Systems IV, ed. {R.~A.~Shaw, H.~E.~Payne, \& J.~J.~E.~Hayes},
  433

\bibitem[{{Skrutskie} {et~al.}(2006){Skrutskie}, {Cutri}, {Stiening},
  {Weinberg}, {Schneider}, {Carpenter}, {Beichman}, {Capps}, {Chester},
  {Elias}, {Huchra}, {Liebert}, {Lonsdale}, {Monet}, {Price}, {Seitzer},
  {Jarrett}, {Kirkpatrick}, {Gizis}, {Howard}, {Evans}, {Fowler}, {Fullmer},
  {Hurt}, {Light}, {Kopan}, {Marsh}, {McCallon}, {Tam}, {Van Dyk}, \&
  {Wheelock}}]{skrutskie2006}
{Skrutskie}, M.~F., {Cutri}, R.~M., {Stiening}, R., {et~al.} 2006, \aj, 131,
  1163

\bibitem[{{Stutz} {et~al.}(2009){Stutz}, {Rieke}, {Bieging}, {Balog},
  {Heitsch}, {Kang}, {Peters}, {Shirley}, \& {Werner}}]{stutz2009}
{Stutz}, A.~M., {Rieke}, G.~H., {Bieging}, J.~H., {et~al.} 2009, \apj, 707, 137

\bibitem[{{Testi} {et~al.}(2014){Testi}, {Birnstiel}, {Ricci}, {Andrews},
  {Blum}, {Carpenter}, {Dominik}, {Isella}, {Natta}, {Williams}, \&
  {Wilner}}]{testi2014}
{Testi}, L., {Birnstiel}, T., {Ricci}, L., {et~al.} 2014, Protostars and
  Planets VI, 339

\bibitem[{{Tobin} {et~al.}(2012{\natexlab{a}}){Tobin}, {Hartmann}, {Bergin},
  {Chiang}, {Looney}, {Chandler}, {Maret}, \& {Heitsch}}]{tobin2012a}
{Tobin}, J.~J., {Hartmann}, L., {Bergin}, E., {et~al.} 2012{\natexlab{a}},
  \apj, 748, 16

\bibitem[{{Tobin} {et~al.}(2008){Tobin}, {Hartmann}, {Calvet}, \&
  {D'Alessio}}]{tobin2008}
{Tobin}, J.~J., {Hartmann}, L., {Calvet}, N., \& {D'Alessio}, P. 2008, \apj,
  679, 1364

\bibitem[{{Tobin} {et~al.}(2012{\natexlab{b}}){Tobin}, {Hartmann}, {Chiang},
  {Wilner}, {Looney}, {Loinard}, {Calvet}, \& {D'Alessio}}]{tobin2012}
{Tobin}, J.~J., {Hartmann}, L., {Chiang}, H.-F., {et~al.} 2012{\natexlab{b}},
  \nat, 492, 83

\bibitem[{{Tobin} {et~al.}(2010{\natexlab{a}}){Tobin}, {Hartmann}, \&
  {Loinard}}]{tobin2010b}
{Tobin}, J.~J., {Hartmann}, L., \& {Loinard}, L. 2010{\natexlab{a}}, \apjl,
  722, L12

\bibitem[{{Tobin} {et~al.}(2010{\natexlab{b}}){Tobin}, {Hartmann}, {Looney}, \&
  {Chiang}}]{tobin2010a}
{Tobin}, J.~J., {Hartmann}, L., {Looney}, L.~W., \& {Chiang}, H.
  2010{\natexlab{b}}, \apj, 712, 1010

\bibitem[{{Tobin} {et~al.}(2010{\natexlab{c}}){Tobin}, {Hartmann}, {Looney}, \&
  {Chiang}}]{tobin2010}
{Tobin}, J.~J., {Hartmann}, L., {Looney}, L.~W., \& {Chiang}, H.-F.
  2010{\natexlab{c}}, \apj, 712, 1010

\bibitem[{{Tobin} {et~al.}(2011){Tobin}, {Hartmann}, {Chiang}, {Looney},
  {Bergin}, {Chandler}, {Masqu{\'e}}, {Maret}, \& {Heitsch}}]{tobin2011}
{Tobin}, J.~J., {Hartmann}, L., {Chiang}, H.-F., {et~al.} 2011, \apj, 740, 45

\bibitem[{{Tobin} {et~al.}(2013){Tobin}, {Bergin}, {Hartmann}, {Lee}, {Maret},
  {Myers}, {Looney}, {Chiang}, \& {Friesen}}]{tobin2013}
{Tobin}, J.~J., {Bergin}, E.~A., {Hartmann}, L., {et~al.} 2013, \apj, 765, 18

\bibitem[{{Tobin} {et~al.}(2015){Tobin}, {Dunham}, {Looney}, {Li}, {Chandler},
  {Segura-Cox}, {Sadavoy}, {Melis}, {Harris}, {Perez}, {Kratter},
  {J{\o}rgensen}, {Plunkett}, \& {Hull}}]{tobin2015}
{Tobin}, J.~J., {Dunham}, M.~M., {Looney}, L.~W., {et~al.} 2015, \apj, 798, 61

\bibitem[{{Tobin} {et~al.}(2016){Tobin}, {Kratter}, {Persson}, {Looney},
  {Dunham}, {Segura-Cox}, {Li}, {Chandler}, {Sadavoy}, {Harris}, {Melis}, \&
  {P{\'e}rez}}]{tobin2016}
{Tobin}, J.~J., {Kratter}, K.~M., {Persson}, M.~V., {et~al.} 2016, \nat, 538,
  483

\bibitem[{{Ulrich}(1976)}]{ulrich1976}
{Ulrich}, R.~K. 1976, \apj, 210, 377

\bibitem[{{van der Bliek} {et~al.}(2004){van der Bliek}, {Norman}, {Blum},
  {Probst}, {Montane}, {Galvez}, {Warner}, {Tighe}, {Delgado}, \&
  {Martinez}}]{vanderbliek2003}
{van der Bliek}, N.~S., {Norman}, D., {Blum}, R.~D., {et~al.} 2004, in
  \procspie, Vol. 5492, Ground-based Instrumentation for Astronomy, ed.
  A.~F.~M. {Moorwood} \& M.~{Iye}, 1582--1589

\bibitem[{{Vorobyov} \& {Basu}(2006)}]{vorobyov2006}
{Vorobyov}, E.~I., \& {Basu}, S. 2006, \apj, 650, 956

\bibitem[{{Whitney} {et~al.}(2003{\natexlab{a}}){Whitney}, {Wood}, {Bjorkman},
  \& {Cohen}}]{whitney2003b}
{Whitney}, B.~A., {Wood}, K., {Bjorkman}, J.~E., \& {Cohen}, M.
  2003{\natexlab{a}}, \apj, 598, 1079

\bibitem[{{Whitney} {et~al.}(2003{\natexlab{b}}){Whitney}, {Wood}, {Bjorkman},
  \& {Wolff}}]{whitney2003}
{Whitney}, B.~A., {Wood}, K., {Bjorkman}, J.~E., \& {Wolff}, M.~J.
  2003{\natexlab{b}}, \apj, 591, 1049

\bibitem[{{Whitney} {et~al.}(2003{\natexlab{c}}){Whitney}, {Wood}, {Bjorkman},
  \& {Wolff}}]{whitney2003a}
---. 2003{\natexlab{c}}, \apj, 591, 1049

\bibitem[{{Woermann} {et~al.}(2001){Woermann}, {Gaylard}, \&
  {Otrupcek}}]{woermann2001}
{Woermann}, B., {Gaylard}, M.~J., \& {Otrupcek}, R. 2001, \mnras, 325, 1213

\bibitem[{{Yen} {et~al.}(2015){Yen}, {Koch}, {Takakuwa}, {Ho}, {Ohashi}, \&
  {Tang}}]{yen2015}
{Yen}, H.-W., {Koch}, P.~M., {Takakuwa}, S., {et~al.} 2015, \apj, 799, 193

\bibitem[{{Yen} {et~al.}(2017){Yen}, {Koch}, {Takakuwa}, {Krasnopolsky},
  {Ohashi}, \& {Aso}}]{yen2017}
---. 2017, \apj, 834, 178

\bibitem[{{Yen} {et~al.}(2013){Yen}, {Takakuwa}, {Ohashi}, \& {Ho}}]{yen2013}
{Yen}, H.-W., {Takakuwa}, S., {Ohashi}, N., \& {Ho}, P.~T.~P. 2013, \apj, 772,
  22

\bibitem[{{Zhao} {et~al.}(2017){Zhao}, {Caselli}, {Li}, \&
  {Krasnopolsky}}]{zhao2017}
{Zhao}, B., {Caselli}, P., {Li}, Z.-Y., \& {Krasnopolsky}, R. 2017, ArXiv
  e-prints, arXiv:1706.06504

\bibitem[{{Zinnecker} {et~al.}(1998){Zinnecker}, {McCaughrean}, \&
  {Rayner}}]{zinnecker1998}
{Zinnecker}, H., {McCaughrean}, M.~J., \& {Rayner}, J.~T. 1998, \nat, 394, 862

\end{thebibliography}
